\documentclass{aastex63}
\usepackage{graphicx}
\usepackage{newtxtext,newtxmath}
\usepackage{psfig}
\usepackage{natbib}
\usepackage{ulem,color}

\def\asec{\ifmmode ^{\prime\prime}\else$^{\prime\prime}$\fi}

\def\msun{\hbox{M$_{\odot}$}}

\def\msunyr{\mbox{\,${\rm M_{\odot}\, yr^{-1}}$}}
\def\mdot{\dot M}
\def\Mdot{\dot M}

\def\degs{\ifmmode ^{\circ}\else$^{\circ}$\fi}
\def\amin{\ifmmode ^{\prime}\else$^{\prime}$\fi}
\def\asec{\ifmmode ^{\prime\prime}\else$^{\prime\prime}$\fi}
\def\farcs{\hbox{$.\!\!^{\prime\prime}$}}  
\def\h{\hbox{$^{\rm h}$}}
\def\m{\hbox{$^{\rm m}$}}
\def\s{\hbox{$^{\rm s}$}}

\def\degs{\ifmmode ^{\circ}\else$^{\circ}$\fi}
\def\amin{\ifmmode ^{\prime}\else$^{\prime}$\fi}

\def\EE#1{\times 10^{#1}}

\def\cm{\mbox{\,cm}}

\def\cm3{\mbox{\,cm$^{-3}$}}
\def\kms{\mbox{\,km~s$^{-1}$}}
\def\cms{\mbox{\,cm~s$^{-1}$}}

\def\kms{\mbox{\,km s$^{-1}$}}

\def\s{\mbox{\,s}}

\def\lsim{\!\!\!\phantom{\le}\smash{\buildrel{}\over
 {\lower2.5dd\hbox{$\buildrel{\lower2dd\hbox{$\displaystyle<$}}\over
                                 \sim$}}}\,\,}
\def\gsim{\!\!\!\phantom{\ge}\smash{\buildrel{}\over
{\lower2.5dd\hbox{$\buildrel{\lower2dd\hbox{$\displaystyle>$}}\over
                               \sim$}}}\,\,}
\received{December 30, 2019}
\revised{January 15, 2020}
\accepted{January 16, 2020}
\shorttitle{Radio Observations of Nearby Type Ia Supernovae}
\shortauthors{Lundqvist et al.}

\begin{document}

\title{The Deepest Radio Observations of Nearby Type IA Supernovae: Constraining Progenitor Types and Optimizing Future Surveys}

\correspondingauthor{Peter Lundqvist}
\email{peter@astro.su.se}

\author{Peter Lundqvist}
\affiliation{Department of Astronomy, AlbaNova University Center, Stockholm University, SE-10691 Stockholm, Sweden}
\affiliation{The Oskar Klein Centre, AlbaNova, SE-10691 Stockholm, Sweden}

\author{Esha Kundu}
\affiliation{Department of Astronomy, AlbaNova University Center, Stockholm University, SE-10691 Stockholm, Sweden}
\affiliation{The Oskar Klein Centre, AlbaNova, SE-10691 Stockholm, Sweden}
\affiliation{Curtin Institute of Radio Astronomy, Curtin University, GPO Box U1987, Perth WA 6845, Australia}

\author{Miguel~A. P\'erez-Torres}
\affiliation{Instituto de Astrof\'isica de Andaluc\'ia, Glorieta de las Astronom\'ia, s/n, E-18008 Granada, Spain}
\affiliation{Visiting Scientist: Departamento de F\'isica Teorica, Facultad de Ciencias, Universidad de Zaragoza, Spain}

\author{Stuart~D. Ryder}
\affiliation{Dept. of Physics and Astronomy, Macquarie University, Sydney NSW 2109, Australia}
\affiliation{Australian Astronomical Observatory, 105 Delhi Rd, North Ryde, NSW 2113, Australia}

\author{Claes-Ingvar Bj\"ornsson}
\affiliation{Department of Astronomy, AlbaNova University Center, Stockholm University, SE-10691 Stockholm, Sweden}

\author{Javier Moldon}
\affiliation{Jodrell Bank Centre for Astrophysics, School of Physics and Astronomy, University of Manchester, M13 9PL, UK}

\author{Megan~K. Argo}
\affiliation{Jodrell Bank Centre for Astrophysics, School of Physics and Astronomy, University of Manchester, M13 9PL, UK}
\affiliation{Jeremiah Horrocks Institute, University of Central Lancashire, Preston PR1 2HE, UK}

\author{Robert~J. Beswick}
\affiliation{Jodrell Bank Centre for Astrophysics, School of Physics and Astronomy, University of Manchester, M13 9PL, UK}

\author{Antxon Alberdi}
\affiliation{Instituto de Astrof\'isica de Andaluc\'ia, Glorieta de las Astronom\'ia, s/n, E-18008 Granada, Spain}

\author{Erik~C. Kool}
\affiliation{Department of Astronomy, AlbaNova University Center, Stockholm University, SE-10691 Stockholm, Sweden}
\affiliation{The Oskar Klein Centre, AlbaNova, SE-10691 Stockholm, Sweden}
\affiliation{Dept. of Physics and Astronomy, Macquarie University, Sydney NSW 2109, Australia}
\affiliation{Australian Astronomical Observatory, 105 Delhi Rd, North Ryde, NSW 2113, Australia}



\begin{abstract}
We report deep radio observations of nearby Type Ia Supernovae (SNe Ia) with the electronic Multi-Element Radio Linked Interferometer Net-work (e-MERLIN), and the Australia Telescope Compact Array (ATCA). No detections were made.  With standard assumptions for the energy densities of relativistic electrons going into a power-law energy distribution, and the magnetic field strength  ($\epsilon_{\rm e} = \epsilon_{\rm B} = 0.1$), we arrive at the upper limits on mass-loss rate for the progenitor system of SN~2013dy~(2016coj, 2018gv, 2018pv, 2019np), to be $\dot{M} \lsim 12~(2.8,1.3, 2.1, 1.7) \EE{-8} \msunyr (v_w/100 \kms)$, where $v_w$ is the wind speed of the mass loss. To  SNe~2016coj, 2018gv, 2018pv and 2019np we add radio data for 17 other nearby SNe~Ia, and model their non-detections.  With the same model as described, all 21 SNe~Ia have $\dot{M} \lsim 4 \EE{-8} \msunyr (v_w/100 \kms)$. We compare those limits with the expected mass loss rates in different single-degenerate progenitor scenarios. We also discuss how information on $\epsilon_{\rm rel}$ and $\epsilon_{\rm B}$ can be obtained from late observations of SNe~Ia and the youngest SN~Ia remnant detected in radio, G1.9+0.3, as well as stripped-envelope core-collapse SNe. We highlight SN~2011dh, and argue for $\epsilon_{\rm e} \approx 0.1$ and $\epsilon_{\rm B} \approx 0.0033$. Finally, we discuss strategies to observe at  radio frequencies to maximize the chance of detection, given the time since explosion, the distance to the supernova and the telescope sensitivity.
\end{abstract}

\keywords{Supernovae: individual - objects: SN\,2006X, SN\,2011dh, SN\,2011fe, SN\,2012cg, SN\,2013dy, SN\,2014J, SN\,2016coj, SN\,2018gv, SN\,2018pv, SN\,2019np, Supernova remnants: individual - G1.9+0.3}


\section{Introduction}
\label{sec:intro}
Type Ia supernovae (SNe~Ia) have proven to be of fundamental 
importance as cosmological distance indicators \citep[e.g.,][]{rie98,per99}. Even so, we are still ignorant
regarding what progenitor scenario is the correct one for the
majority of SNe~Ia. This compromises their use for 
precision cosmology. In addition, they are key players in the
chemical evolution of galaxies, but not knowing the details of
progenitor evolution, the explosion and the nucleosynthesis, 
means we do not fully understand the timescale over 
which SNe~Ia turn on, adding uncertainty to models for the chemical
enrichment in the Universe.

It is a generally accepted fact that SNe~Ia are thermonuclear explosions of 
white dwarfs (WDs) \citep{hoy60}. There are mainly two competing 
classes of models leading to a SN Ia thermonuclear explosion. One is 
the double-degenerate (DD) model where two WDs merge and explode
\citep[e.g.,][]{tut79,iben84,Webb84,tho11,mao14}. The other is the  single-degenerate (SD) model, where 
the companion is a non-degenerate
star \citep[e.g.,][]{whe73,nom82,wang18}. Here the WD accretes matter 
from the companion until it undergoes unstable runaway nuclear
burning. A branch of these models is the so-called spun-up/spun-down
super-Chandrasekhar mass WDs \citep{dis11,jus11} where mass transfer 
is no longer active at the time of explosion.

One way to discriminate among different progenitor models 
of SNe~Ia is to obtain information about the circumstellar medium of
the exploding star. In scenarios with mass transfer from a
non-degenerate companion, non-conservative mass transfer will give rise
to a circumstellar medium \citep[see, e.g.,][]{bra95} with a structure
that depends on the mass-loss history of the system. When the SN 
ejecta are expelled into this medium, a shock is bound to form,
resulting in radio and X-ray emission \citep{che82b}. In the DD
scenario, the surrounding medium is likely to be of interstellar
origin, and also in the SD spun-up/spun-down scenario one can expect 
a low-density medium in the vicinity of the progenitor. In these
two scenarios, essentially no radio or X-ray emission is expected.

Several early attempts were made to detect 
radio \citep[e.g.,][]{panagia06,han11} 
and X-ray \citep[e.g.,][]{hug07,rus12} emission from SNe~Ia. These
searches were hampered by their limited sensitivity and some inadequate
assumptions for the modeling. 
 The situation improved with the
emergence of the very nearby SNe~2011fe and 2014J, for which
sensitive observations could be made. Using methods of interpretation
incorporated from stripped-envelope SNe, upper limits on the
mass-loss rate from the progenitor systems have been obtained.
Radio and X-ray limits for these two SNe~Ia 
suggest $\mdot \lsim 10^{-9}$ \msunyr\ \citep{cho12,cho16,per14}, 
and $\mdot \lsim 2\times10^{-9}$ \msunyr\ \citep{mar12,mar14},
respectively, assuming a wind velocity of 100~km~s$^{-1}$.  
In addition to this, \citet{cho16} have compiled a very
comprehensive list of deep observations with the Jansky Very Large
Array (JVLA) of nearby SNe~Ia. Here we report five more SNe~Ia to
add to this list from our ongoing programs on the electronic Multi-Element
Radio Linked Interferometer Net-work (e-MERLIN), and the Australia
Telescope Compact Array (ATCA), namely SNe~2013dy, 2016coj, 2018gv, 
2018pv and 2019np. Like in previous attempts, for other SNe~Ia, we do not
detect these five SNe in the radio.

The non-detections of radio and X-ray emission from SNe~Ia have 
added to a growing consensus that SNe~Ia mainly stem from DD 
explosions \citep[e.g.,][]{mao14}, but a potential problem is that no obvious
candidate system with double WDs has ever been identified \citep{reb19}. 
This, however, seems consistent with the intrinsic faintness of these 
objects. For potential SD progenitors, one should not disregard the 
SD spun-up/spun-down scenario, and/or that the generation of radio and
X-ray emission could be less efficient than hitherto assumed. Also,
there is in fact evidence of circumstellar material from time-varying
absorption features in the Na~I~D line for some SNe~Ia
\citep{pat07,sim09}. 
The exact location of this material
is still debated, and there is no support for the idea
that shells around SNe~Ia, which give rise to dust scattering, are 
of circumstellar origin \citep{bul18}. 

There is a subset of SNe~Ia which indeed
show clear evidence of circumstellar interaction, the first ones
being SNe~2002ic \citep{ham03} and 2005gj \citep{ald06}, and the first case 
with both circumstellar interaction and time-varying narrow absorption 
lines was PTF~11kx \citep{dil12}. The most recently reported circumstellar
interaction examples are SNe~2015cp
\citep{gra18} and 2018fhw \citep{Val19}. All these show Balmer line emission, so their
progenitor systems are with little doubt of SD origin. \citet{gra18}
estimate that $< 6$\% of all SNe~Ia have circumstellar shells within
$3\EE{17}$~cm from the exploding star. Due to selection effects, this fraction 
could be even smaller.

At some time after the explosion, the SN will turn on as a radio
source, even if one has to wait until the supernova remnant (SNR)
stage. A local example is G1.9+0.3, and there is also a hint that SN~1885A
in Andromeda may now be visible at radio wavelengths \citep{sar17}. 
We discuss the information
we can gain from these to use in models for young SNe~Ia.

Here we first describe the radio observations of SNe~2013dy, 
2016coj, 2018gv, 2018pv and 2019np (Section ~\ref{sec:obs}), and in 
Section~\ref{sec:model} we discuss the model we are using to interpret the
observations. In Section~\ref{sec:results} we summarize the results
for the five SNe. Then, in Section~\ref{sec:discuss}, we choose
the 21 best observed SNe~Ia in radio, along with the youngest local 
SN~Ia remnant seen in radio (SNRG1.9+0.3),
 to draw some conclusions about 
what radio observations of SNe~Ia can actually constrain in terms of the
nature of the progenitor system. We also discuss optimal strategies for observing
SNe~Ia in terms of time since explosion, radio frequency and sensitivity. Finally, we wrap 
up the paper in Section~\ref{sec:conclusions} with our main conclusions.

\section{Observations and data reduction}
\label{sec:obs}
The data for our observations of the five nearby SNe~Ia 2013dy, 2016coj, 2018gv, 2018pv and 2019np are collected in Tables~\ref{tab:RadioLog} and \ref{tab:RadioLog2}. Here we describe these observations.

\subsection{SN~2013dy}
We observed SN~2013dy in the nearby (D=13.7 Mpc) galaxy NGC~7250 with the electronic Multi Element Radio 
Interferometric Network (e-MERLIN) \citep{per13}. SN2013dy was discovered on 2013 Julyt 10.45 UT 
\citep{casper13,zheng13}, and  our radio observations were carried out during 2013 August 4 - 6, about 
one week after the SN had reached its B-band maximum.  We observed SN~2013dy with e-MERLIN at a central 
frequency of 5.09 GHz, and used a total bandwidth of 512 MHz, which resulted in a synthesized Gaussian beam 
of 0$\farcs$13 x 0$\farcs$11. We centered our observations at the position of the optical discovery,
and followed standard calibration and imaging procedures. We imaged a $20\arcsec \times 20\arcsec$ region 
centered at this position, after having stacked all our data. We found no evidence of radio emission above a 
3$\sigma$ limit of 300~$\mu$Jy\,bm$^{-1}$ in a circular region of 1\arcsec\ in radius, centered at the SN position. 
This value corresponds to an upper limit of the monochromatic 
5.0 GHz luminosity of $6.7\EE{25}$ erg s$^{-1}$ Hz$^{-1}$ (3$\sigma$). 

\begin{table*}
\caption{Parameters of observed Type Ia SNe}
\scalebox{1.0}{
\begin{tabular}{llccccc}
\tableline\tableline
Supernova  & Date of Optical Max & SN Position                    &  Host Galaxy & Host Type & Distance & SN References \cr
           &     (UT)            &  (J2000.0)                     &              &           & (Mpc)    &               \cr
\tableline
%
 %
SN 2013dy   &  2013 Jul 27.71   & 22:18:17.60, +40:34:09.6   & NGC 7250    &   Sdm     &   13.7   &   1         \cr
SN 2016coj  &  2016 Jun 08.35   & 12:08:06.80, +65:10:38.2   & NGC 4125    &   E pec   &   20.1   &   2         \cr  
SN 2018gv   &  2018 Feb 3       & 08:05:34.61, $-$11:26:16.3 & NGC 2525    &   SB(s)c  &   16.8   & 3           \cr
SN 2018pv   &  2018 Feb 16      & 11:52:55.70, +36:59:11.6    & NGC 3941    &   SB(s)   &   13.1  &   4         \cr
SN 2019np   &  2019 Jan 26      & 10:29:21.96, +29.30.38.4      & NGC 3254    &   SA(s)bc   &   22  &   5         \cr
\tableline
\end{tabular}
}
\tablecomments{The columns starting from left to right are as follows: Supernova name; Date of optical maximum;  Coordinates of the supernova in the optical; Host galaxy; Host galaxy type; Distance; Light curve maximum estimate: (1) \citet{zheng13}, (2) \citet{zheng17}, (3) Chen et al. (in preparation), (4) Subo Dong (private communication), (5) Subo Dong and Nancy Elias-Rosa (private communication)}
 \label{tab:RadioLog}
\end{table*}

\subsection{SN 2016coj}
We observed SN~2016coj in the nearby (D=20.1 Mpc) galaxy NGC 4125 with e-MERLIN on 2016 May 28.18 UT(MJD 57536.18) \citep{per16} . Our observations were carried out on 2016 June 3-4, one week after the SN discovery and about one week before reaching its V-band maximum  \citep{zheng16,zheng17}.   e-MERLIN observed at a central frequency of 1.51 GHz and used a total bandwidth of 512 MHz, which resulted in a synthesized Gaussian beam of $0\farcs13 \times 0\farcs12$. We centered our observations at the position of the optical discovery, 
and imaged a $16\arcsec \times 16\arcsec$ region centered at this position. We found no evidence of radio emission in the region of SN~2016coj down to a 3$\sigma$ limit of 126 $\mu$Jy\,bm$^{-1}$, which corresponds to an upper limit of the monochromatic 1.51 GHz luminosity of $6.1\EE{25}$ erg s$^{-1}$ Hz$^{-1}$ (3$\sigma$). 

\begin{table*}
\caption{Observations of studied supernovae}
\scalebox{0.8}{
\begin{tabular}{llcccccrrc}
\tableline\tableline
SN Name & Observation & Facility &  Central & Time Since  & Flux    &  Luminosity   & $\dot M/v_w$     & $n_0$ & References       \cr
        &  Date       &          &  Freq.   & Explosion   & Density (1$\sigma$) &  Upper Limit (3$\sigma$)  & Upper Limit       & Upper Limit &  \cr
        &        &          &    &      &  &   &   &   & \cr  
        &    (UT)      &          &  (GHz)    &   (Days)    & ($\mu$Jy) & ($10^{25}$ erg s$^{-1}$ Hz$^{-1}$) & ($\frac {10^{-8}\msunyr}{100 \kms}$)  & ($\cm3$) & \cr  
\tableline
%
 
 SN 2013dy  & 2013 Aug 4-6    & e-MERLIN  &  5.09  & 26          & 100    &  6.74   & 12 & 300  &    1,2 \cr
 SN 2016coj & 2016 June 3.42  & e-MERLIN &  1.51  & 11          & 42     &  6.09   & 2.8     & 240  &    1,3 \cr
            & 2016 June 3.86  & AMI     &  15.0  & 11          & 101    &  14.6   & 18     & 2300 &    4   \cr
            & 2016 June 5.89  & AMI     &  15.0  & 13          &  74    &  10.7   & 17     & 1600 &    4   \cr
            & 2016 June 9.76  & AMI     &  15.0  & 17          &  52    &  7.54   & 18     & 1000  &    4   \cr
            & 2016 June 11.07 & JVLA     &  2.7   & 18          &  20    &  2.95   & 3.3     & 120  &    4   \cr
            &                 & JVLA     &  4.5   & 18          &  20    &  2.90   & 4.5     & 180  &    4   \cr
            &                 & JVLA     &  7.4   & 18          &  16    &  2.32   & 5.4     & 220  &    4   \cr
            &                 & JVLA     &  8.5   & 18          &  17    &  2.42   & 6.1     & 260  &    4   \cr
            &                 & JVLA     &  10.9  & 18          &  18    &  2.56   & 7.4     & 330  &    4   \cr
            &                 & JVLA     &  13.5  & 18          &  13    &  1.93   & 7.1     & 310  &    4   \cr
            &                 & JVLA     &  16.5  & 18          &  16    &  2.36   & 9.3     & 420  &    4   \cr
            & 2016 June 11.81 & AMI     &  15.0  & 19          &  65    &  9.47   & 23     & 1200  &    4   \cr
            & 2016 June 12.81 & AMI     &  15.0  & 20          &  65    &  12.8  & 30     & 1400  &    4   \cr
            & 2016 June 13.81 & AMI     &  15.0  & 21          &  65    &  10.4  & 27     & 1100  &    4   \cr
SN 2018gv   & 2018 Jan 18.6   & ATCA    &  5.5   &  6          &  40    &  4.05   & 2.3     & 610  &    1,5 \cr
            &                 & ATCA    &  9.0   &  6          &  10    &  1.01   & 1.3     & 300  &    1,5 \cr
SN 2018pv   & 2018 Feb 9.23   & e-MERLIN &  5.1   & 14          &  19    &  1.18   & 2.1     & 120  &    1,6 \cr
SN 2019np  & 2019 Jan 11.97   & MeerKAT &  1.28   & 7         &  19    &  3.30   & (1.8)\tablenotemark{a}     & 220  &    7 \cr
                     & 2019 Jan 14.81   & e-MERLIN &  1.51   & 10          &  22    &  3.82   & 1.7     & 160  &    1,8 \cr
 \tableline
\end{tabular}
}
\tablecomments{The columns starting from left to right are as follows: Supernova name; Starting observing time, UT; Observational Facility; 
Central frequency in GHz; Mean observing epoch (in days since explosion); 1$\sigma$ flux density upper limits, in $\mu$Jy; 
 The corresponding 3$\sigma$ spectral luminosity in units of $10^{25}$\,erg\,s$^{-1}$\,Hz$^{-1}$; Inferred 3$\sigma$ upper limit to the  mass-loss rate in units of $10^{-8} M_\odot\,{\rm yr}^{-1}$, for an  assumed wind velocity of $100 \kms$. (The values for $\dot M$ are for $\epsilon_{\rm B} = \epsilon_{\rm e} = 0.1$.); Inferred 3$\sigma$ upper limit to the circumstellar density for a constant density medium, in units of $\cm3$. References are in the last column: (1) This paper; (2) \citet{per13}, (3) \citet{per16}, (4) \citet{mol16} and data tabulated on the web as described in Section 2.2, (5) \citet{ryd18}, (6) \citet{per18}, (7) \citet{hey19}, (8) \citet{per19}. 
 $^{\rm a}$No solution exists (cf. Figure~\ref{fig:Time_Mdot}).}
 \label{tab:RadioLog2}
\end{table*}

In our analysis we also include data from AMI and the Jansky VLA (JVLA). In addition to what is reported in \citet{mol16}, further data are tabulated here\footnote{https://4pisky.org/atel-sn2016coj-20160627/}. These data cover epochs from 2016 June 3.86 to 2016 June 13.81, estimated to correspond to 15$-$25 days after explosion (cf. Table~\ref{tab:RadioLog2}).

\subsection{SN 2018gv}
We used the Australia Telescope Compact Array (ATCA) at 5.5 and 9.0 GHz with 2~GHz bandwidths on 2018 Jan~18.6~UT to observe SN~2018gv \citep{ryd18} situated in the galaxy NGC 2525. This SN was discovered on 2018 Jan~15.681~UT by Koichi Itagaki (TNS discovery report \#16498), and identified as a SN Ia by \citet{bufano18} and \citet{sieb18}. The observations and data reduction followed the same procedures as outlined for SN~2011hs by \citet{bufano14}. 
No radio emission was detected down to 3$\sigma$ upper limits of 120 $\mu$Jy \,bm$^{-1}$ at 5.5 GHz, and 30 $\mu$Jy\,bm$^{-1}$ at 9.0 GHz. The  total on-source time at each frequency was of 6.8 hr. Adopting the host galaxy distance from \citet{tully13} of 16.8~Mpc, this implies an upper limit on the 9.0 GHz luminosity of $1.0\EE{25}$ erg s$^{-1}$ Hz$^{-1}$ (3$\sigma$), and four times higher at 5.5 GHz. 

\subsection{SN 2018pv}
We observed the SN~Ia 2018pv with e-MERLIN at 5.1 GHz on  2018 February 3.63 UT (MJD 58153.13) in the nearby ($z=0.0031$) galaxy NGC 3941 (Tsuboi, TNS discovery report \#16800). A spectrum on 8.78 February 2018 (MJD 58158.78) confirmed the SN as a Type Ia event a few days before maximum \citep{yama18}. Our observations \citep{per18} were carried out on $9-10$ February 2018 UT (MJD 58159.08), six days after the SN discovery. We centered our observations at the position of the optical discovery 
(cf. Table~\ref{tab:RadioLog}). We found no evidence of radio emission in a circular region of $4\farcs0$ diameter surrounding SN~2018pv, down to a 3$\sigma$ upper limit of 57.6 $\mu$Jy\,bm$^{-1}$. For an assumed distance of 13.1 Mpc, the corresponding upper limit on the monochromatic 5.1 GHz luminosity is of $1.2\EE{25}$ erg s$^{-1}$ Hz$^{-1}$ (3$\sigma$).

\subsection{SN 2019np}
We observed the SN~Ia 2019np with e-MERLIN between 2019 January 14.81 and 15.46 UT \citep{per19}. SN~2019np was
 discovered on 2019 January 9.67 UT in the nearby ($z=0.00452$) galaxy NGC 3254 (Itagaki, TNS discovery report \#28550),
  and a spectrum on 2019 January 10.83 UT confirmed the SN as a Type Ia event two weeks before maximum (Burke, TNS 
  classification report \#3399). This is probably a lower limit since B-band maximum appears to have occurred around 2019  January 26 (S. Dong and N. Elias-Rosa, private communication). Our observations were 
  thus carried out 5 days after the SN discovery, and $t \lsim 10$ days after the SN explosion. For a conservative estimate of $t$ we have used 10 days. We observed at a central frequency of 1.51 GHz, with a bandwidth of 512 MHz, and centered our observations at the position of the optical discovery (cf. Table~\ref{tab:RadioLog}). We found no evidence of radio emission in a circular region of $10\farcs0$ diameter surrounding SN~2019np, down to a 3$\sigma$ upper limit of 66 $\mu$Jy\,bm$^{-1}$. For an assumed distance of 22 Mpc, the corresponding upper limit of the monochromatic 1.51 GHz luminosity is of $3.82\EE{25}$ erg s$^{-1}$ Hz$^{-1}$ (3$\sigma$). In our analysis we also include MeerKAT observations, commencing at 2019 January 11.97 UT \citep{hey19}. The total integration lasted 3.25 hours in the frequency band 856 -- 1690 MHz.The observation resulted in a 3$\sigma$ upper limit of 57$\mu$Jy\,bm$^{-1}$ at 1280 MHz, corresponding to $3.30\EE{25}$ erg s$^{-1}$ Hz$^{-1}$ (3$\sigma$). We have used $t =7$ days, but this should be considered as an upper limit on $t$. 

\section{Modelling the radio emission from SNe~Ia}
\label{sec:model}

We now interpret the upper limits on radio emission from the SNe in \S\ref{sec:obs}  within the framework of circumstellar interaction. The supernova shock-wave expands out into its circumstellar gas, and a high-energy density shell forms. Here electrons are accelerated to relativistic speeds and significant magnetic fields are generated. The relativistic electrons radiate synchrotron emission \citep[e.g.,][]{che82b}, which we probe with our radio observations.

We use the same model for the radio emission as in \citet{per14} and \citet{kun17}. In particular, we assume that electrons are accelerated to relativistic energies, with a power law distribution, $dN/dE = N_0E^{-p}$; where $E=\gamma m_ec^2$ is the energy of the electrons and $\gamma$ is the Lorentz factor.
For synchrotron emission, the intensity of optically thin emission $\propto \nu^{-\alpha}$, where $\alpha = (p-1)/2$. As shown for Type Ibc SNe, $\alpha\approx1$ \citep{che06}, and we therefore use $p=3$ as our default value. 
 
\par 
The density of the ambient medium as a function of radial distance, $r$, can be given as $\rho(r) = n_{\rm CSM}(r)\mu$, where 
$n_{\rm CSM}(r)$ and $\mu$ are the particle density and mean atomic weight of the surrounding gas,
respectively. In the case of a constant density medium we put $n_{\rm CSM}(r) = n_0$, and for a wind medium 
$\rho(r) \propto r^{-s}$. For constant $\mdot/v_w$,  where $\dot M$ and $v_w$ are the mass loss rate of the progenitor and the 
velocity at which this mass has been ejected from the system, respectively, $\rho(r) = \dot M/(4\pi r^2 v_w)$. In our models,
we test the two scenarios $s = 0$ and $s = 2$. 

\par 
For the SN ejecta, we resort to two models, also discussed in \citet{kun17}. One is called the N100 model \citep{rop12,sei13}, 
and tests the SD scenario. This is a delayed detonation model where the central region is ignited by 100 sparks. The other is
known as a violent merger model \citep{pak12}, which probes the DD channel. In this, two C/O degenerate stars with masses 
of $1.1~\msun$ and $0.9~\msun$ merge and produce a successful SN explosion. The total masses and asymptotic kinetic 
energies of the ejecta for N100 and the violent merger model are $1.4~\msun$, $1.95~\msun$, and $1.45\EE{51}$ erg and 
$1.7\EE{51}$ erg, respectively.

For both these models, the density profiles of the ejecta are given by the numerical simulations up to around a velocity of $2.5 \times 10^{4}$ $\kms$. Therefore, for the extreme outer part of the exploded WD a power law density structure is considered, i.e., $\rho_{ej} \propto r^{-n}$. In this study we have assumed $n = 13$ \citep[see][for a discussion on $n$]{kun17}.     

\par 
The interaction of the supersonic SN ejecta with the almost stationary ambient medium creates two shock waves, 
known as forward and reverse shocks. In the shocked gas encapsulated by these shocks, relativistic particles are accelerated in the presence of magnetic fields, and synchrotron radiation is emitted at radio wavelengths. We assume that the radio emission comes from a spherical homogeneous shell, and that the evolution of this shell is described by a self-similar solution \citep{che82a}.

\par 
For a polytropic gas with $\gamma = 5/3$, the compression of the gas across the strong shock is $\eta = 4$, and the post-shock 
thermal energy density is $u_{\rm th} = 9/8 \rho(r) v_s^2(r)$, where $v_s(r)$ is the velocity of the forward shock at a given 
distance $r$. We assume that fractions of the thermal energy, $\epsilon_{\rm e} = u_{\rm e}/u_{\rm th}$ and   
$\epsilon_{\rm B} = u_{\rm B}/u_{\rm th}$, go into the energy densities of electrons ($u_{\rm e}$) and magnetic 
fields ($u_{\rm B} = B^2/(8\pi)$), respectively. Here $B$ is the magnetic field strength. We assume that in the post-shock region, 
all electrons get accelerated. However, with time only a fraction of the electrons, represented by $\epsilon_{\rm rel}$, remains 
relativistic with energy $E > m_ec^2$, where $m_e$ and $c$ are the mass of electron and velocity of light, respectively.  These 
relativistic electrons are the ones which give rise to radio emission. 

Following \citet{per14} we have in our models considered 
synchrotron self-absorption (SSA) as the sole absorption mechanism of this radiation \citep[see also the discussion in][]{kun17}. 
In the optically thin regime, from a shell of radius $r_s$ and thickness of $\Delta r$, the luminosity can be written as follows

\begin{equation}
	L_{\nu, {\rm thin}}=\frac{8 \pi^{2} kT_{\rm bright} \vartheta_{\nu} r_s^2} {c^2 f\left(\frac{{\nu}_{\rm peak}}{{\nu}_{\rm abs}}\right)}
	                 \nu_{{\rm abs},0}^{(p+3)/2} \nu^{-(p-1)/2},
\label{eq:Luminosity}
\end{equation}
with 
\begin{equation}
	\nu_{{\rm abs},0}= \left(2 \Delta r ~ \varkappa(p) ~N_0 ~ B^{(p+2)/2}\right)^{2/(p+4)}
\label{eq:nuabs}
\end{equation}
and
\begin{equation}
	f (x) = x^{1/2} \left[1 - {\rm exp} \left( - x^{-\left(p+4\right)/2} \right) \right]
\label{eq:fx}
\end{equation}
\citep{per14,kun17}, where $k$ and $T_{\rm bright}$ represent the Boltzmann constant and the brightness temperature, respectively. In this work it is assumed that $T_{\rm bright} = 5 \times 10^{10}$ K, which is the same value as that considered in \citet{per14} and \citet{kun17}. Note that
$T_{\rm bright}$ is defined from the intensity at $\nu_{{\rm abs},0}$ \citep[cf.][]{bjo14}. $\vartheta_{\nu} = \frac{L_{\nu}}{4 \pi^2 r_{s}^{2} I_{\nu}(0)}$, with $I_{\nu}(0)$ being the intensity of radiation received from the equatorial plane of the SN, i.e., from that part of the shell for which path length is equal to $\Delta r$ along the line of sight. $\varkappa(p)$ and $B$ are the SSA coefficient and magnetic field strength in the post-shock region, respectively. For $n = 13$ and $p = 3$, the optically thin luminosity can be written for a constant density medium, $s = 0$, as 

\begin{equation}
	L_{\nu, {\rm thin}} \propto T_{\rm bright} ~ \epsilon_{\rm e}^{1.71} ~ \epsilon_{\rm B}^{1.07}  n_{0}^{1.28} ~ t^{0.91},
\label{eq:Lum11}
\end{equation}
and for a wind medium with $s = 2$ as
\begin{equation}
	L_{\nu, {\rm thin}} \propto T_{\rm bright} ~ \epsilon_{\rm e}^{1.71} ~ \epsilon_{\rm B}^{1.07}  \left(\dot M / v_w\right)^{1.51} ~ t^{-1.42}.
\label{eq:Lum12}
\end{equation}

\section{Results}
\label{sec:results}
\subsection{Modeling the data for our sample}
\label{sec:modelingsample}
Radio emission from SNe Ia is attenuated by free-free absorption (FFA)
in the external unshocked circumstellar medium, and by SSA. In early analyses of SNe~Ia \citep[e.g.,][]{panagia06,han11}, 
FFA was 
considered to dominate the absorption. However, more recent papers
\citep{cho12,hor12,per14,cho16,kun17}, conclude that FFA is insignificant. As
discussed in \citet{per14}, the free-free optical
depth, $\tau_{\rm ff}$, for a fully ionized wind at $10^4$~K and 
moving at 
$v_w = 100~\kms$, is $\tau_{\rm ff} \sim 10^{-4} \lambda^{2} (\Mdot/10^{-7} \msunyr)^2 (r_s/10^{15} {\rm cm})^{-3}$, 
where $\lambda$ is in cm. From our calculations, using the N100 model,
the shock radius is at $\sim 10^{15}$~cm already at $\sim 2$ days 
for $\Mdot= 10^{-7} (v_w/100 \kms)$\msunyr, which means that
$\tau_{\rm ff} \sim 3\EE{-3} (\Mdot/10^{-7} \msunyr)^2$ at 
5.5 GHz at that epoch. Considering that X-ray non-detections for 
SNe~2011fe and 2014J \citep{mar12,mar14} have put limits on 
$\mdot/v_w$ of order $10^{-9} \msunyr$ for $v_w = 100 \kms$ for these
SNe~Ia, it is not a bold assumption that FFA can be neglected for
normal SNe~Ia. \citet{hor12} used a similar argument to
dismiss free-free absorption in their analysis of radio emission from 
SN~2011fe. In what follows, we only discuss frequencies higher than 1 GHz, and concentrate
on $\Mdot \lsim 10^{-7}(v_w/100 \kms)$\msunyr and $t \gsim 2$ 
days, and therefore only consider SSA.

\begin{figure}
\centering 
\includegraphics[width= 12cm,angle=0]{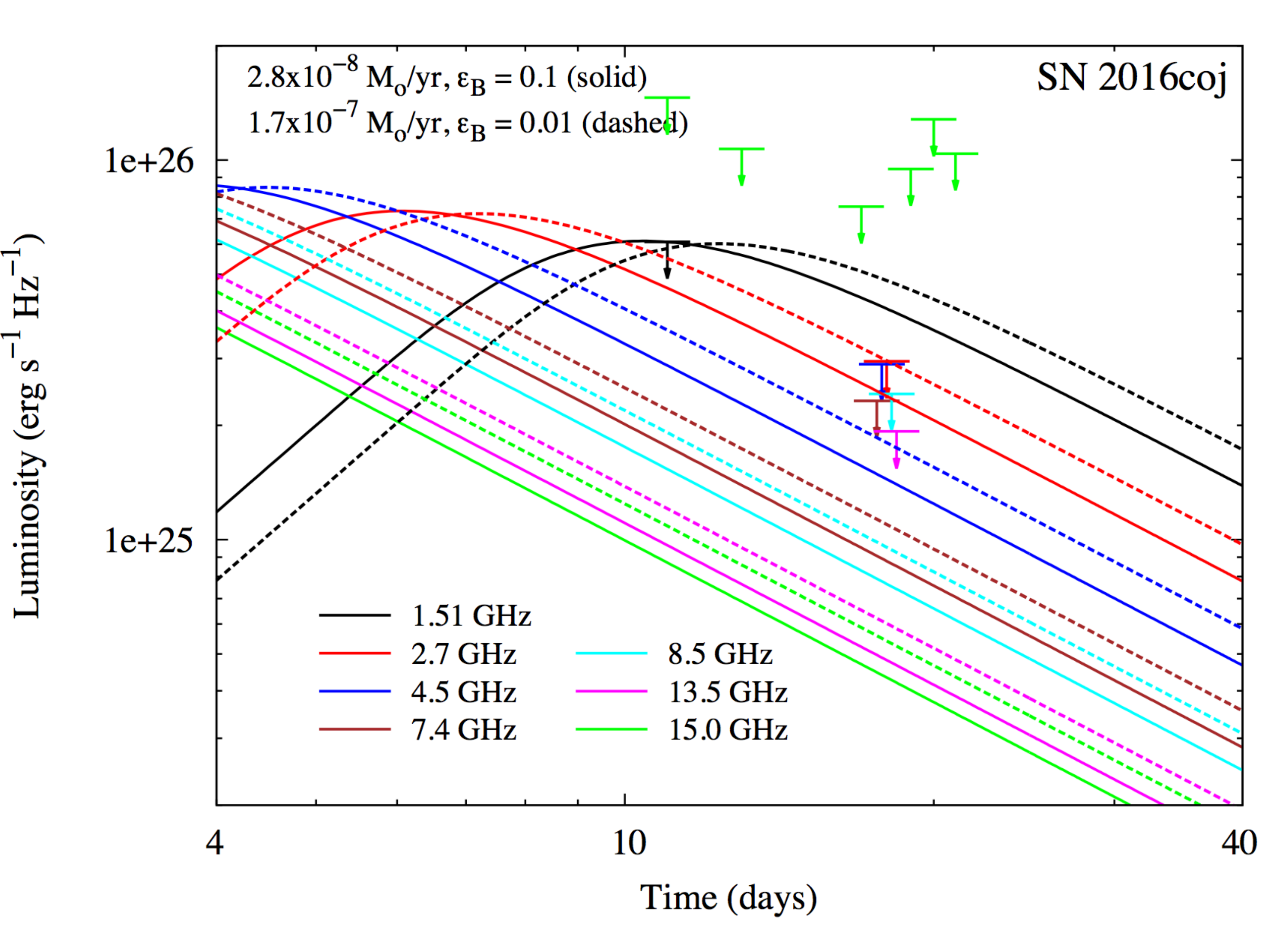}
\caption{
Radio data for SN~2016coj (see Table~\ref{tab:RadioLog2}) together
with models at various frequencies for an $s=2$ wind. Models use 
$\dot{M} = 2.8~(17) \EE{-8} \msunyr~(v_w/100 \kms)$
for $\epsilon_{\rm B} = 0.1~(0.01)$, with solid lines being for
$\epsilon_{\rm B} = 0.1$. Common parameters in both models
are $\epsilon_{\rm e} = 0.1$, $T_{\rm bright} = 5\EE{10}~{\rm K}$,
$n=13$ and $v_w = 100~\kms$. Constraining observations are those at 1.51 GHz on 
day 11, and at 2.7 GHz on day 18. Observations at different frequencies on day 18 
have been shifted in steps of 0.2 days between $17.6-18.4$ days to disentangle 
the data.
}
\label{fig:Lcurves_s2_oursample1}
\end{figure}

\begin{figure}
\centering
\includegraphics[width=12cm,angle=0]{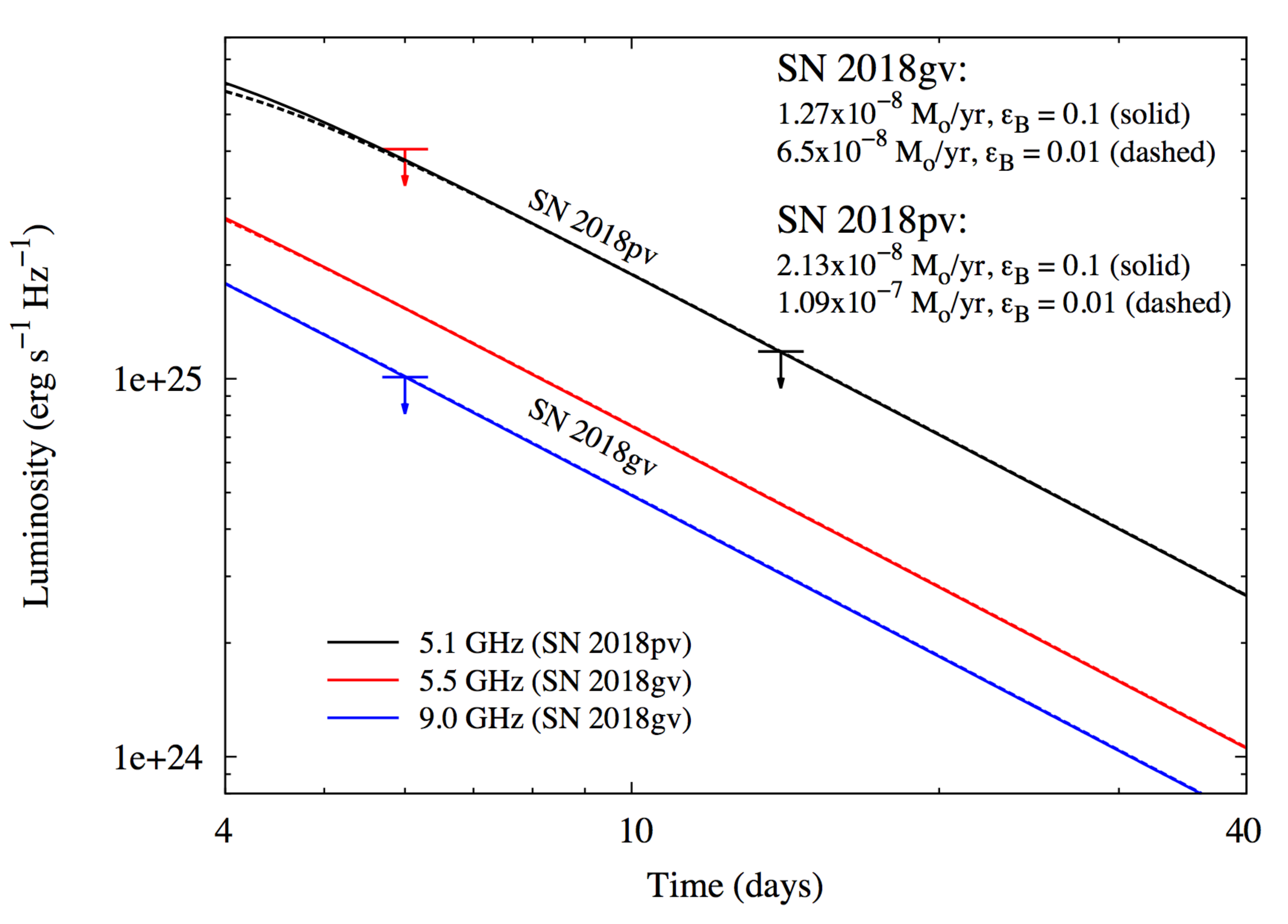}
\caption{
Radio data for SNe~2018gv and 2018pv (see Table~\ref{tab:RadioLog2}) 
together with models at various frequencies for an $s=2$ wind. Common 
model parameters are $\epsilon_{\rm e} = 0.1$, $T_{\rm bright} =
5\EE{10}~{\rm K}$, $n=13$ and $v_w = 100 \kms$. Solid lines are for
$\epsilon_{\rm B} = 0.1$ and dashed for $\epsilon_{\rm B} = 0.01$. 
Note that dashed and solid lines overlap for SN~2018pv. The values
for $\dot{M}$ in the different models are described in the figure.
The constraining observations are at 5.1 GHz for SN~2018pv and 
9.0 GHz for SN~2018gv.
}
\label{fig:Lcurves_s2_oursample2}
\end{figure}

\begin{figure}
\centering
\includegraphics[width=12cm,angle=0]{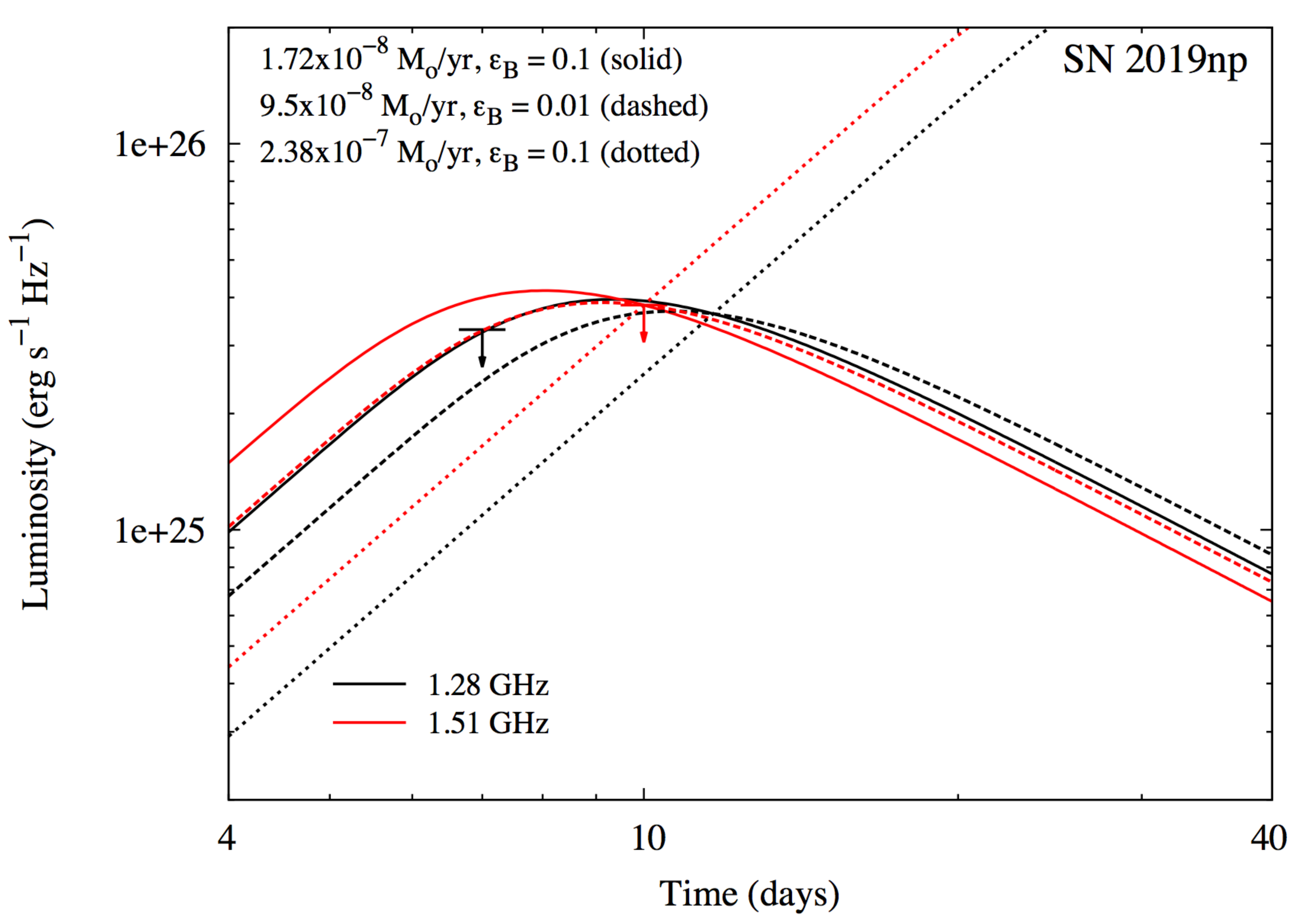}
\caption{
Radio data for SN~2019np (see Table~\ref{tab:RadioLog2}) 
together with models at two frequencies for an $s=2$ wind. Common 
model parameters are $\epsilon_{\rm e} = 0.1$, $T_{\rm bright} =
5\EE{10}~{\rm K}$, $n=13$ and $v_w = 100 \kms$. Solid and dotted lines are for
$\epsilon_{\rm B} = 0.1$ and dashed for $\epsilon_{\rm B} = 0.01$. 
The values for $\dot{M}$ in the different models are described in the figure. For 
$\epsilon_{\rm B} = 0.1$ and $v_w = 100 \kms$, 
$1.7\times10^{-8}\msunyr \leq {\dot M} \leq 2.4\times10^{-7}\msunyr$ is
ruled out from the observation at 1.51 GHz. 
}
\label{fig:Lcurves_s2_oursample3}

\end{figure}

We have used the model in Section~\ref{sec:model} to calculate the
expected emission from a circumstellar medium created by a wind 
(the $s=2$ case), and for a constant-density medium (the $s=0$ case).
Expressions for epochs when
SSA is negligible are given by Equations~\ref{eq:Lum11} and
~\ref{eq:Lum12}. These expressions can be used to study the
dependence between the various parameters, and are in most cases
sufficient in order to estimate $\Mdot/v_w$ and $n_0$. However, 
SSA can be important at very early epochs and especially at low
frequencies, so the expressions for optically thin synchrotron
emission may underestimate $\Mdot/v_w$ and $n_0$. As discussed in 
Section~\ref{sec:model}, our models do include SSA.

\subsubsection{The constant density case, $s=0$.}
\label{sec:constant}
We have used the merger model and methods discussed 
in Section~\ref{sec:model} to estimate $n_0$ for SNe~2013dy, 
2016coj, 2018gv, 2018pv and 2019np. As shown in Table~\ref{tab:RadioLog2}, 
the lowest limit on $n_0$ for those SNe~Ia is 
$n_0 \gsim 120~\cm3$ (for SNe~2016coj and 
2018pv), assuming $\epsilon_{\rm B} = \epsilon_{\rm e} = 0.1$. This 
is significantly higher than the density expected in the DD scenario, 
which is that of the ISM, i.e., $\lsim 1 \cm3$. This shows that 
early radio observations of SNe~Ia do not provide stringent limits on $n_0$,
unless they are significantly  closer than 20 Mpc.

\begin{figure*}
\centering
\includegraphics[width=18.8cm,angle=0]{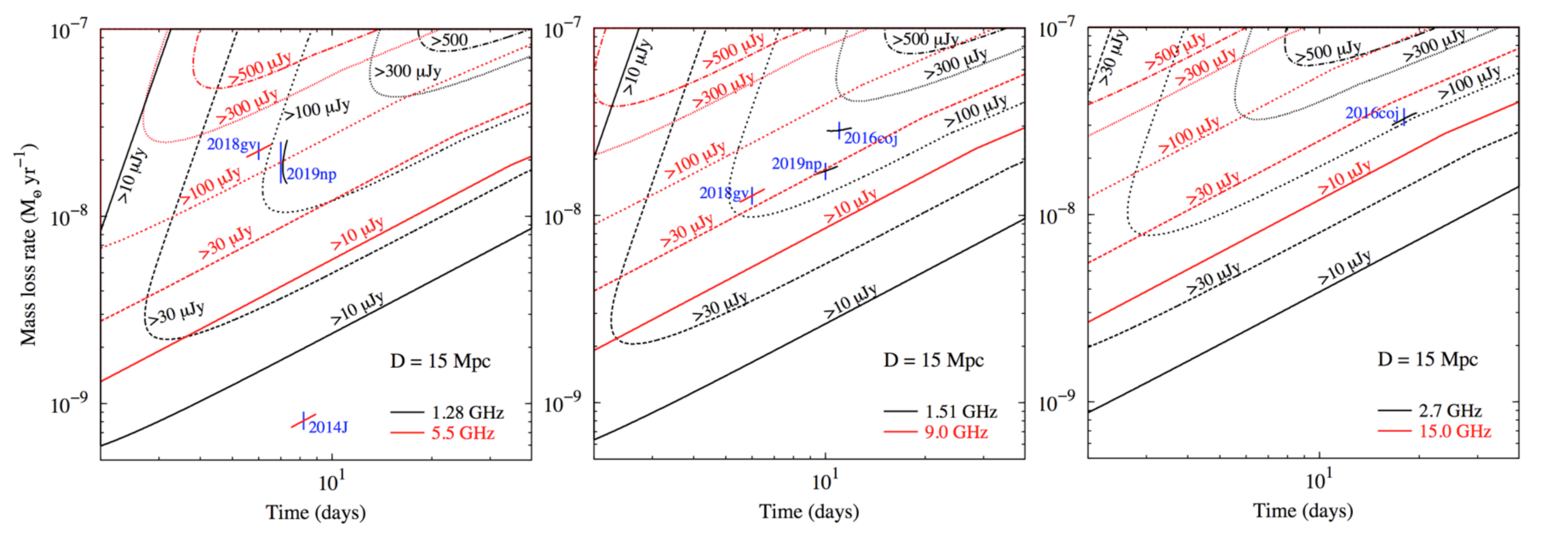}
\caption{
Wind density (in mass-loss rate per year) probed by radio observations as a function of time since SN~Ia explosion.
Solution curves for given fluxes in $\mu$Jy are drawn for six observing frequencies: 1.28 GHz and 5.5 GHz (left panel), 1.51 
and 9.0 GHz (middle panel), and 2.7 GHz and 15.0 GHz (right panel). The assumed distance to the supernova is 15 Mpc, and 
solution curves are drawn for 10~$\mu$Jy (solid lines), 30~$\mu$Jy (dashed), 100~$\mu$Jy (dotted) , 300~$\mu$Jy 
(finely dotted)  and 500~$\mu$Jy (dashed-dotted). Part of solution 
curves, for the times since explosion, are also drawn for the 3$\sigma$ upper limits of SNe~2014J, 2016coj, 2018gv 
and 2019np, where the flux limits (cf. Tables~\ref{tab:RadioLog2} and \ref{tab:RadioLog3}) have been adjusted to the distance
 of 15 Mpc. The parameters used to calculate the solution curves are $T_{\rm bright} = 5\times10^{10}$~K,  
$v_w = 100~\kms$, $\epsilon_{\rm B} = \epsilon_{\rm e} = 0.1$, and the N100 model with $n=13$. Solution curves, and
vertical tick marks marking the time since explosion, cross for the $\dot M/v_w$ values in 
Tables~\ref{tab:RadioLog2} and \ref{tab:RadioLog3}. Note the effect of synchrotron self-absorption at the lowest frequencies 
and the highest mass-loss rates, which means that there is a minimum time since explosion when the supernova can be 
detected for a given flux limit and observing frequency.
}
\label{fig:Time_Mdot}
\end{figure*}

As the radio luminosity in the $s=0$ case is expected to increase 
with time \citep[e.g.,][]{cho12,per14,kun17}, radio observations
at late epochs constrain $n_0$ better \citep[see, e.g.,][]{cho16}. 
For events nearby enough, like SNe~2011fe and 2014J, tight limits on
both $n_0$, and the microphysics parameters $\epsilon_{\rm B}$ and
$\epsilon_{\rm e}$ can be obtained \citep[][see also Section~ \ref{sec:eps}]{kun17}. Modeling data from
the epochs 1468 days and 410 days, and assuming
$\epsilon_{\rm B} = \epsilon_{\rm e} = 0.1$, \citet{kun17} find 
$n_0 \lsim 0.3~\cm3$ for both SNe~2011fe and 2014J, respectively. 
According to \citet{cho16}, limits for other SNe~Ia do not come 
close to these numbers, the best cases being SNe~1985A and 2012cg.
\citet{cho16} report $n_0 \lsim 13~(10) \cm3$ for SN~1985A
(SN~2012cg)
observed at 1.4~(5.9) GHz observations at 315~(216) days. For the 
sake of completeness, we have recalculated the corresponding values using 
our models and methods in Section~\ref{sec:model}, together with the data in \citet{cho16} and using 
$\epsilon_{\rm B} = \epsilon_{\rm e} = 0.1$. We find 
$n_0 \lsim 12~(8) \cm3$ for SN~1985A~(SN~2012cg), which is close 
to the numbers of \citet{cho16}.

\subsubsection{The wind case, $s=2$.}
\label{sec:wind}
While limits on $n_0$ in the $s=0$ for young SNe~Ia case are of 
limited value, except for SNe~2011fe and 2014J, early radio
observations can be 
used to constrain $\dot{M}/v_{w}$ in the $s=2$ case with some
stringency.
As shown in Table~\ref{tab:RadioLog2}, deep limits on
$\dot{M}/v_{w}$ are obtained for SNe~2016coj, 2018gv, 2018pv
and 2019np.
For $\epsilon_{\rm B} = \epsilon_{\rm e} = 0.1$, and using the 
N100 explosion model with $n=13$, we find upper limits of 
$\dot{M} \lsim 2.8~(1.3, 2.1, 1.7) \EE{-8} \msunyr (v_w/100 \kms)$, 
for these three SNe, respectively. The limit on 
$\dot{M}/v_{w}$ for SN~2013dy is about an order of magnitude 
larger.

We show modeled light curves for SN~2016coj in
Figure~\ref{fig:Lcurves_s2_oursample1}, for SNe 2018gv and 2018pv
in Figure~\ref{fig:Lcurves_s2_oursample2}, and for SN~2019np in Figure~\ref{fig:Lcurves_s2_oursample3}. 
All models use $\epsilon_{\rm e} = 0.1$, $T_{\rm bright} = 5\EE{10}~{\rm K}$ and 
$n=13$, and we show results for both $\epsilon_{\rm B} = 0.01$ and
$\epsilon_{\rm B} = 0.1$. For SN~2016coj, the most constraining data
are from the e-MERLIN 1.51 GHz observations on day 11 (cf. Table~\ref{tab:RadioLog2}), 
but the JVLA data at 2.7 GHz also provide
stringent constraints. In particular, for $\epsilon_{\rm B} = 0.01$, SSA
is important at 1.51 GHz, while the optically thin
2.7 GHz emission not only serves as an independent check, but also sets a more 
stringent limit on $\dot{M}/v_w$. The mass-loss rate limit for the $\epsilon_{\rm B} = 0.01$ case is
$\dot{M} \lsim 1.7~ \EE{-7} \msunyr (v_w/100 \kms)$, i.e., 
almost an order of magnitude higher than for $\epsilon_{\rm B} = 0.1$.

In the models for SNe 2018gv and 2018pv, SSA does not play a role for
the $5-9$ GHz light curves in Figure~\ref{fig:Lcurves_s2_oursample2},
not even for the models with $\epsilon_{\rm B} = 0.01$. Models with
$\epsilon_{\rm B} = 0.1$ and $\epsilon_{\rm B} = 0.01$ line up on
top of each other, just by changing $\dot{M}/v_{w}$ by a factor of 
$10^{0.71} \approx 5.1$, as expected from Equation~\ref{eq:Lum12}
for optically thin synchrotron radiation. The corresponding factor
is larger ($\approx 6.7$) for the marginally optically thick
situation at 1.51 GHz in Figure~\ref{fig:Lcurves_s2_oursample1}.
The limits on 
$\dot{M}/(v_w/100 \kms)$ for $\epsilon_{\rm B} = 0.01$ and
$\epsilon_{\rm B} = 0.1$ for SNe~2018gv and 2018pv are shown in
Figure~\ref{fig:Lcurves_s2_oursample2}. 

\begin{table*}
\caption{Type Ia SNe with the most constraining data for a wind-like circumstellar scenario.}
\scalebox{0.85}{
\begin{tabular}{llcccccrr}
\tableline\tableline
SN Name & Host, and & Distance &  Central & Time Since  & Flux    &  Luminosity   & $\dot M/v_w$   & Reference         \cr
        &  Host Type       &          &  Freq.   & Explosion   & Density (1$\sigma$) &  Upper Limit (3$\sigma$)  & Upper Limit       &   \cr
        &        &          &    &      &  &   &   &    \cr  
        &       &    (Mpc)   &  (GHz)    &   (Days)    & ($\mu$Jy) & ($10^{25}$ erg s$^{-1}$ Hz$^{-1}$) & ($\frac {10^{-8}\msunyr}{100 \kms}$)  & \cr  
\tableline
%
 SN 1989B   & NGC 3627, SAB(s)b    & 10      &  4.8  & 13           & 30    &  1.08   & 1.8      &    1,2  \cr
 SN 1995al  & NGC 3021, SA(rs)bc   & 27      &  1.4  & 17           & 80    &  6.98   & 4.0      &    1,2  \cr
 SN 2006X   & NGC 4321. SAB(s)bc   & 17      &  8.4  & 5.9          & 18    &  1.87   & 1.8      &    1  \cr
 SN 2010fz  & NGC 2967, SA(s)c     & 31      &  6.0  & 8.9          & 9     &  3.11   & 3.0      &    1  \cr
 SN 2011at  & PGC 26905, SB(s)d    & 25      &  5.9  & 20.2         & 5     &  1.12   & 3.2      &    1  \cr
 SN 2011by  & NGC 3972, SA(s)bc    & 20      &  5.9  & 8.1          & 3     &  0.48   & 0.78     &    1  \cr
 SN 2011dm  & UGC 11861, SABdm     & 20      &  5.9  & 14.1         & 5     &  0.72   & 1.7      &    1  \cr
 SN 2011ek  & NGC 918, SAB(rs)c    & 18      &  5.9  & 7.4          & 5     &  0.58   & 0.81     &    1  \cr
 SN 2011fe  & M101, SAB(rs)cd      & 6.4     &  5.9  & 2.1          & 5.8   &  0.118  & 0.087    &    1,3\cr
 SN 2011iv  & NGC 1404, E1         & 19      &  6.8  & 6.2          & 18    &  3.11   & 2.3      &    1 \cr
 SN 2012Z   & NGC 1309, SA(s)bc    & 29      &  5.9  & 7.0          & 7     &  1.81   & 1.6      &    1 \cr
 SN 2012cg  & NGC 4424, SB(s)a     & 15      &  4.1  & 5            & 5     &  0.40   & 0.35     &    1 \cr
 SN 2012cu  & NGC 4772, SA(s)a     & 29      &  5.9  & 15.2         & 5     &  1.61   & 3.1      &    1 \cr
 SN 2012ei  & NGC 5611, S0         & 25      &  5.9  & 16.0         & 6     &  1.35   & 2.9      &    1 \cr
 SN 2012fr  & NGC 1365, SB(s)b     & 18      &  5.9  & 3.9          & 7     &  1.12   & 0.69     &    1 \cr
 SN 2012ht  & NGC 3447, Pec        & 20      &  5.9  & 4.6          & 5     &  0.91   & 0.70     &    1 \cr
 SN 2014J   & M 82, Irr            & 3.4     &  5.5  & 8.2          & 4     &  0.0167 & 0.081    &    1,3\cr
 SN 2016coj & NGC 4125, E pec      & 20.1    &  1.51 & 11           & 42    &  6.09   & 2.8      &    4,5 \cr
 SN 2018gv  & NGC 2525, SB(s)c     & 16.8    &  9.0  &  6           & 10    &  1.01   & 1.3      &    4,6 \cr
 SN 2018pv  & NGC 3941, SB(s)      & 13.1    &  5.1  &  14          & 19    &  1.18   & 2.1      &    4,7 \cr
 SN 2019np   & NGC 3254, SA(s)bc  & 22   &  1.51   & 10          &  22    &  3.82   & 1.7     &    4,8 \cr
\tableline
\end{tabular}
}
\tablecomments{The columns starting from left to right are as follows: Supernova name; Host galaxy and galaxy type; 
Distance in Mpc; Central frequency in GHz; Mean observing epoch (in days since explosion); 1$\sigma$ flux density upper limits, in $\mu$Jy; The corresponding 3$\sigma$ spectral luminosity in units of $10^{25}$\,erg\,s$^{-1}$\,Hz$^{-1}$; Inferred 3$\sigma$ upper limit to the  mass-loss rate in units of $10^{-8} M_\odot\,{\rm yr}^{-1}$, for an  assumed wind velocity of $100 \kms$. (The values for $\dot M$ are for $\epsilon_{\rm B} = \epsilon_{\rm e} = 0.1$.); References are in the last column: (1) \citet{cho16}; (2) \citet{panagia06}; (3) \citet{per14}; (4) This paper; (5) \citet{per16}; (6) \citet{ryd18}; (7) \citet{per18}; (8) \citet{per19}.}
 \label{tab:RadioLog3}
\end{table*}

For SN~2019np, SSA is important at the low frequencies (1.28 GHz and 1.51 GHz) used for observations 
of this supernova (cf. Figure~\ref{fig:Lcurves_s2_oursample3}). For 1.28 GHz at $t=7$ days, the peak luminosity 
for $\epsilon_{\rm B} = 0.1$ 
is $3.25\EE{25}$ erg s$^{-1}$ Hz$^{-1}$, and occurs for $\dot{M} \approx 1.8~ \EE{-8}  (v_w/100 \kms) \msunyr$.  This
1.28 GHz luminosity is lower than the $3\sigma$ limit listed in Table~\ref{tab:RadioLog2}. To highlight this, we have 
put the upper limit on $\dot{M}/v_w$ for 1.28 GHz in Table~\ref{tab:RadioLog2} in parenthesis. 
For 1.51 GHz, at $t=10$ days, the modeled luminosity for $\epsilon_{\rm B} = 0.1$ is higher than the observed 
$3\sigma$ limit for $1.7\times10^{-8}\msunyr \lsim {\dot M} (v_w/100 \kms)^{-1}  \lsim 2.4\times10^{-7}\msunyr$. 
The corresponding limits for $\epsilon_{\rm B} = 0.01$ are 
$9.5\times10^{-8}\msunyr \lsim {\dot M} (v_w/100 \kms)^{-1} \lsim 5.1\times10^{-7}\msunyr$.
For ${\dot M} (v_w/100 \kms)^{-1} \gsim 2.4~(5.1)\times10^{-7}\msunyr$ and $\epsilon_{\rm B} = 0.1~(0.01)$
SSA mutes the modeled 1.51 GHz luminosity so it becomes lower than the observed $3\sigma$ luminosity limit. 
In Table~\ref{tab:RadioLog2}, and in the following, we have, however, treated $1.7~ \EE{-8} \msunyr (v_w/100 \kms)$ as
a true upper limit for $\epsilon_{\rm B} = 0.1$.

Figure~\ref{fig:Time_Mdot} illustrates the relevance of SSA in probing ${\dot M}/v_w$ from SN Ia observations.
We show, for a putative SN~Ia at a distance of  $D=15$~Mpc, which minimum value of ${\dot M}/v_w$ can be probed, 
given the observing frequency, time since explosion and the flux limit.  We have rescaled the flux density levels for
the SNe marked in the figure to correspond to $D=15$~Mpc. Solution curves for a given flux density level, and vertical 
tick marks marking the time since explosion, overlap for the $\dot M/v_w$ values tabulated in Tables~\ref{tab:RadioLog2} 
and \ref{tab:RadioLog3}. SSA attenuates the flux densities so that there is a minimum time since explosion when the supernova 
can be detected for a given flux limit and observing frequency. For earlier times, SSA is so large that observations cannot 
constrain $\dot M/v_w$. In particular, there is no solution corresponding to the flux limit of the 1.28 GHz observations at 
$t=7$~days for SN~2019np. This is also highlighted in Table~\ref{tab:RadioLog2}, where $\dot M/v_w$ for the closest 
distance between the solution curve and the vertical line marking time since explosion in the panel, has  been put in parenthesis.
The situation is different for 1.51 GHz at 10 days (middle panel; see also Table~\ref{tab:RadioLog3}) can. Figure~\ref{fig:Time_Mdot} provides a useful tool for selecting radio telescope facility and observing
frequency for a newly detected SN~Ia. For very young SNe (i.e., a few days old), the very lowest frequencies 
($\lsim 2$~GHz) should be avoided, unless one can expect a 3$\sigma$ flux limit which is
$\lsim 10~(D/15~{\rm Mpc})^{-2}~\mu$Jy. For a five day old SN~Ia, the corresponding flux limit is 
$\lsim 30~(D/15~{\rm Mpc})^{-2}~\mu$Jy. 

\section{Discussion}
\label{sec:discuss}
\subsection{Comparison to previous studies}
\label{sec:comparison}
As discussed in Section~\ref{sec:constant}, early radio data
are often not useful to probe the $s=0$ scenario, and in the following we will
mainly concentrate on the $s=2$ scenario. 
To put things in perspective, we have in
Table~\ref{tab:RadioLog3} compiled all SNe~Ia with the most
constraining radio data for that scenario. Our four best cases,
SNe~2016coj, 2018gv, 2018pv and SN~2019np, are the four most recent in
this sample of 21 SNe~Ia. To form this sample we have 
added to our SNe the ones with the lowest limits on $\dot{M}/v_{w}$ in the
compilation of \citet{cho16}. In
Table~\ref{tab:RadioLog3} we list
upper limits on $\dot{M}/v_{w}$ using the same model as in 
Section~\ref{sec:wind} with $\epsilon_{\rm B} = 0.1$. According
to such an estimate, no SN in the sample has 
$\dot{M} \gsim 4.0 \EE{-8} (v_w/100 \kms) \msunyr $.
Seven SNe have 
$\dot{M} \lsim 1.0 \EE{-8}  (v_w/100 \kms) \msunyr$, 
and they are SNe~2011by, 2011ek, 2011fe, 2012cg, 2012fr, 
2012hr and 2014J. SNe~2011fe and 2014J have limits as 
low as $\dot{M} \lsim 9 \EE{-10}(v_w/100 \kms) \msunyr$.

The limits on $\dot{M}/v_{w}$ in Table~\ref{tab:RadioLog3} (and used 
throughout this paper) were
derived using the same distances to the SNe as in Section~\ref{sec:obs} 
and \citet{cho16}

\begin{figure*}
\centering
\includegraphics[width=18.9cm,angle=0]{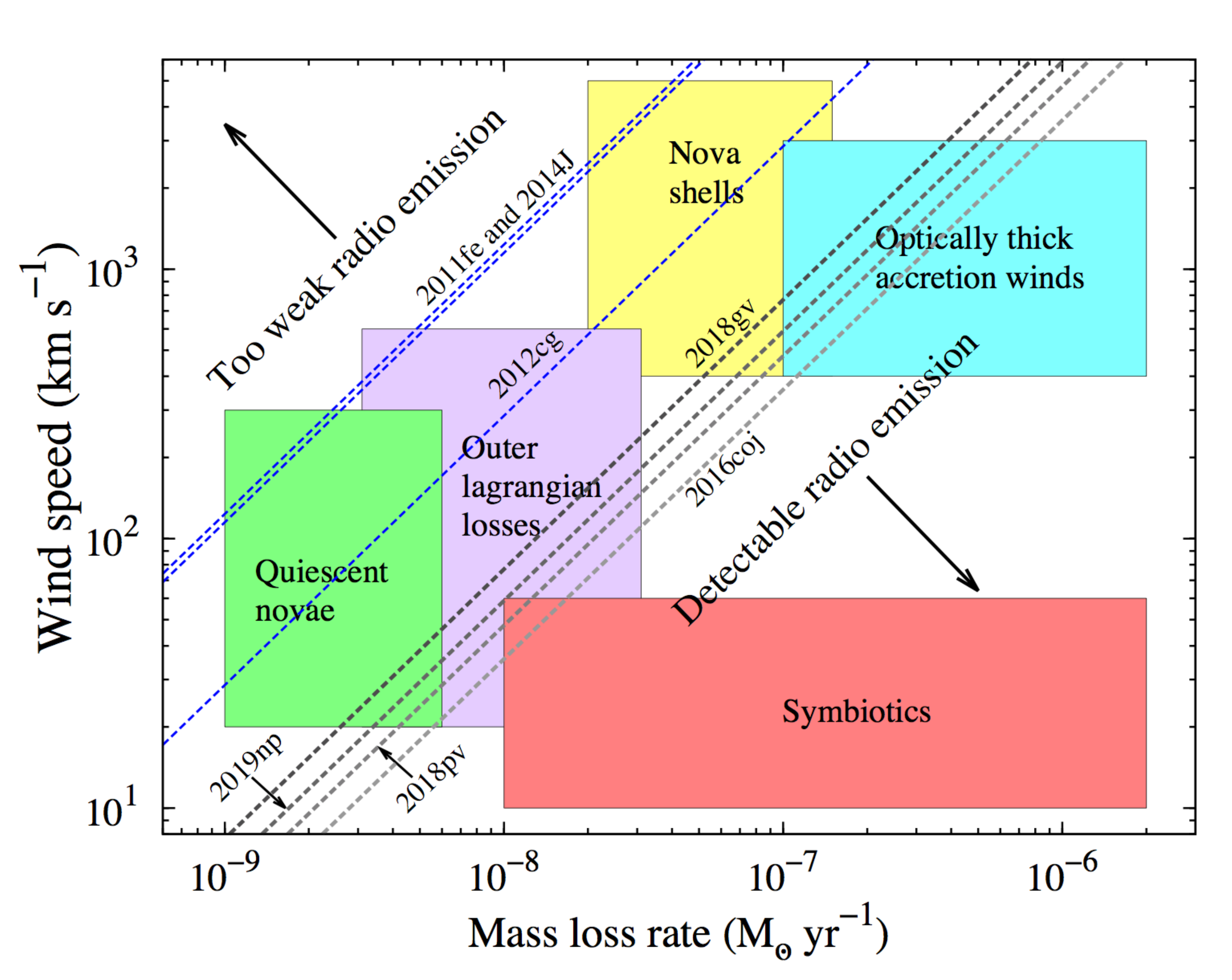}
\caption{
Constraints on the parameter space (wind speed vs. mass-loss
rate) for single degenerate scenarios for seven SNe~Ia. The
progenitor scenarios discussed in \S\ref{sec:sd} are plotted as
schematic zones, following \citet{cho12} and \citet{per14}.
3$\sigma$ limits on $\mdot/v_{w}$ from
Table~\ref{tab:RadioLog3} are marked by dashed lines,
assuming $\epsilon_{\rm B} = \epsilon_{\rm e} = 0.1$, $n=13$,
$T_{\rm bright} = 5\EE{10}$~K, and the N100 explosion model. 
For the parameters in the upper left corner, i.e., low $\mdot$
and high $v_{w}$, the radio emission in $s=2$ scenario is too
weak to be detected, even for events like SNe~2011fe and 2014J.
The opposite is true for the lower right part of the figure.
In particular, the 3$\sigma$ limits disfavor symbiotics as
a likely progenitor channel. Not included in the figure is
the spun-up/spun-down progenitor scenario \citep{dis11,jus11}, 
as this predicts a constant density.
For SNe~2011fe and SN 2014J, only a small part of the parameter
space for likely SD progenitors are possible.
}
\label{fig:mdot-vwind}
\end{figure*}

\subsection{Possible single-degenerate progenitor systems}
\label{sec:sd}
There are several possible SD scenarios, and all (except the
so-called spun-up/spun-down super-Chandrasekhar mass scenario, see
below) are characterized by a mass-loss rate and wind speed of the 
circumstellar gas expelled from the progenitor system.
The expected mass loss rate from the progenitor
system, in decreasing order, includes symbiotic systems, WDs 
with steady nuclear burning, and recurrent novae. We have 
marked areas in Figure~\ref{fig:mdot-vwind} (showing $\mdot$
versus $v_{w}$) where possible SD progenitor systems reside.
We have also marked (dashed lines) 3$\sigma$ limits on
$\mdot/v_{w}$ from Table~\ref{tab:RadioLog3} for seven of the
tabulated SNe, assuming 
$\epsilon_{\rm B} = \epsilon_{\rm e} = 0.1$, $n=13$, 
$s=2$, $T_{\rm bright} = 5\EE{10}$~K, and the N100 explosion
model. Areas in Figure~\ref{fig:mdot-vwind} for the possible 
SD progenitor systems, lying below, and to the right of the
3$\sigma$ limit dashed lines, are ruled out. 

In symbiotic systems (red region in Figure \ref{fig:mdot-vwind}), 
the WD accretes mass from a giant star \citep{hachisu99}, 
but the WD loses some of this  matter at rates 
of $\dot{M} \gtrsim 10^{-8}$\msunyr and
velocity $v_w\approx 30~\kms$. From Figure~\ref{fig:mdot-vwind}
it is clear that this scenario is ruled out for all SNe
in Table~\ref{tab:RadioLog3} with 
$\dot{M} \leq 1.7 \EE{-8} \msunyr (v_w/(100~\kms))$,
including our observed cases SNe~2018gv and 2019np. 
This conclusion, however, rests on $\epsilon_{\rm B} = \epsilon_{\rm e} = 0.1$, 
which is uncertain (cf. Section~\ref{sec:eps}).

Circumstellar medium can also be created during Roche-lobe overflow
from a main sequence, subgiant, helium, or giant star onto the WD.
The expected rate is $3.1\times 10^{-7} \msunyr \lesssim \dot{M}_{\rm acc} 
\lesssim 6.7\times 10^{-7} \msunyr$ \citep{nomoto07}. At those
accretion rates, the WD experiences steady nuclear burning
\citep{shen07}. Assuming an efficiency of 99\%, the mass-loss rate from
the system is $3.1\times 10^{-9} ~\msunyr  \lesssim \dot{M} \lesssim 6.7\times 10^{-9} ~\msunyr$. 
Typical speeds of the gas in the CSM are
$100~\kms \lesssim v_w \lesssim~3000~\kms$. The lower part of the
range is for steady nuclear burning. The highest speeds are relevant
for systems with the highest accretion rates. Of particular interest
is the speed for those systems with the lowest mass-loss rates, and
they lose mass through the outer Lagrangian points at speeds up to
$\sim 600~\kms$. We have marked this region in purple (`Outer
lagrangian losses') in Figure~\ref{fig:mdot-vwind}. With
$\epsilon_{\rm B} = \epsilon_{\rm e} = 0.1$,
systems of this sort are ruled out for SNe~2011fe and 2014J, and 
partly for SN~2012cg, but not for the other SNe.

If the accretion rate is higher, i.e., 
$\dot{M}_{\rm acc} \sim 6\times 10^{-7}~M_{\odot}~\rm yr^{-1}$, 
winds around the WD are 
likely optically thick, limiting the accretion. Any further 
potential mass transfer will be lost from the system at an
expected wind speed of order $10^3 \kms$ \citep{hachisu99,hachisu08}.
This is marked by the cyan-colored box in
Figure~\ref{fig:mdot-vwind}. Assuming
$\epsilon_{\rm B} = \epsilon_{\rm rel} = 0.1$, SNe~2011fe, 2012cg
and 2014J, do not stem from such a type of progenitor system, while
other SNe marked in the figure could.

Systems giving rise to recurrent novae are other possible SN~Ia
progenitors. These systems have low accretion rate, 
$\dot{M}_{\rm acc}  \approx (1-3) \times10^{-7}~\msunyr$. At nova 
outbursts, they eject shells at speeds of a few $\times~1000 \kms$,
with a time between shell ejections of a few, or several years. 
From Table \ref{tab:RadioLog3}, the radio observations probe
observing times between $2-20$ days. For a model with N100, $s=2$ 
and $\dot{M} = 1.0 \EE{-8} \msunyr (v_w/100 \kms)$, the 
shock in our models reaches $\simeq 1.2 \times 10^{16}$~cm. This
constrains the presence of shells with recurrence times of 
$\lesssim 1.9~(v_{\rm shell}/2000~\kms)^{-1}$~yr. Since the nova
ejection is a transient event, the nova shell will be rather
confined, and the likelihood for a supernova shock being caught
while interacting with a nova shell for the first $\sim 20$ days 
is small \citep[about 30\%, according to][]{cho12}. To estimate
$\dot{M}$ during such a phase, we make use of the fact that models of
recurrent novae predict that $\lsim 15$\% of the accreted material
between nova bursts is ejected \citep{yaron2005,shen2009}. We 
follow \citet{cho12}, and highlight the estimated range for 
$\dot{M}$ and $v_w$ with the yellow box in 
Figure~\ref{fig:mdot-vwind}. Using 
$\epsilon_{\rm B} = \epsilon_{\rm e} = 0.1$, we cannot rule out
nova shells completely for SNe~2011fe and 2014J, and not at all 
for the other SNe.

The final box in Figure~\ref{fig:mdot-vwind} is marked in green, 
and is for novae during the quiescent phase between nova shell
ejections. This is most likely for novae with long recurrence
periods, and thus for those with the lowest accretion rates (i.e., 
$\dot{M}_{\rm acc} \sim 1 \times10^{-7}~\msunyr$). The mass-loss from the
system is in this case, 
$\dot{M} \sim 1 \times10^{-9}\ (\epsilon_{\rm loss}/0.01)~
(v_w/100~\kms)$. 
If $\epsilon_{\rm B} = \epsilon_{\rm e} = 0.1$, the models 
rule out almost completely the scenario with WD accretion during 
the quiescent phase of the star for SNe~2011fe and 2014J, whereas
systems with the highest winds and lowest mass loss rates are 
viable possibilities for the other SNe in Table \ref{tab:RadioLog3}.

For Figure~\ref{fig:mdot-vwind} in general, the parameters in the
upper left corner, i.e., low $\mdot$ and high $v_{w}$, the radio
emission is too weak to be detected for any hitherto observed SN~Ia.
The opposite is true for the lower right part of the figure, for
which all the SNe~Ia in Table \ref{tab:RadioLog3} would have been
detected if $\epsilon_{\rm B} = \epsilon_{\rm e} = 0.1$, and if 
they would have belonged to any of the highlighted progenitor 
scenarios in Figure~\ref{fig:mdot-vwind}. In particular, for
SNe~2011fe and SN 2014J, only a small part of parameter space for
possible SD progenitors is allowed.

\subsection{Microphysics parameters $\epsilon_{\rm B}$ and $\epsilon_{\rm rel}$}
\label{sec:eps}
Progenitor constraints on the SNe in Table \ref{tab:RadioLog3}
were discussed in Section \ref{sec:sd} under the assumption of 
$\epsilon_{\rm B} = \epsilon_{\rm e} = 0.1$. This assumption 
has been used in most previous studies
\citep[e.g.,][]{cho12,cho16,
per14,kun17}, although cases with $\epsilon_{\rm B} = 0.01$ have
also been considered. A more general assumption is that
$\epsilon_{\rm B}$ and $\epsilon_{\rm e}$ 
(and thus $\epsilon_{\rm rel}$) can take any
reasonable value, and this may differ from $\epsilon_{\rm e} = 0.1$,
in combination with $0.01 \leq \epsilon_{\rm B} \leq 0.1$. 

As no SN~Ia has yet been detected in the radio, observational
constraints on $\epsilon_{\rm B}$ and $\epsilon_{\rm rel}$ 
can only be obtained from core-collapse SNe, preferably from
stripped-envelope SNe as they have compact progenitors and fast SN ejecta. 
Assuming 
all non-relativistic electrons go into a power-law distribution
with $\gamma_{\rm min} \geq 1$, \citet{che06} argued for 
$\epsilon_{\rm rel} \geq 0.16~(v_s / 5\EE{4}~\kms)^{-2}$ 
and used $\epsilon_{\rm B} \sim 0.1$.
The question is whether the assumptions going into this 
are general. An example is the early phase radio and X-ray emission
of SN~2011dh. \citet{sod12} modeled this emission under the
assumption of $0.1 \lesssim \epsilon_{\rm rel} \lesssim 0.3$ and
$0.01 \lesssim \epsilon_{\rm B} \lesssim 0.1$, and 
arrived at a wind density characterized by
$\dot{M}\approx 6\times 10^{-5}\,M_{\odot}$\ yr$^{-1}$ (for an
assumed wind velocity of $v_w = 1000~\kms$). This is a considerably
less dense environment than estimated using models for the
thermal X-ray emission from the supernova, at the somewhat
later epoch of $\sim 500$ days, for which \citet{mae14} 
and \citet{kun18} estimate $\dot{M} \sim (2-4)\times
10^{-6}\,M_{\odot}$\ yr$^{-1}$ (for  
$v_w = 10~\kms$), i.e., $\sim 5$ times denser than that in the
analysis of \citet{sod12}. This could signal a decreased 
wind density towards the end of the life of the progenitor, but 
it may also be explained by too high values used by \citet{sod12}
for $\epsilon_{\rm B}$ and $\epsilon_{\rm rel}$.
\citet{kun18} found good solutions to radio
light curves at late epochs of the SNe~IIb 1993J and 2011dh 
using $\epsilon_{\rm B} = \epsilon_{\rm e} = 0.03$, and
$\epsilon_{\rm B} = 0.03$ and  $\epsilon_{\rm e} = 0.04$, 
respectively.

One can also gain information about microphysics parameters from
the youngest SN~Ia remnant detected in radio and X-rays, namely
G1.9+0.3 in the Milky Way \citep{con98,rey08}. Models for its radio
emission, assuming a constant density medium around it, suggest 
the use of $\epsilon_{\rm rel} = 10^{-4}$ and 
$\epsilon_{\rm B} \sim 0.01$ \citep[][see also below]{sar17}. 

\begin{figure}
\centering
\includegraphics[width=12 cm,angle=0]{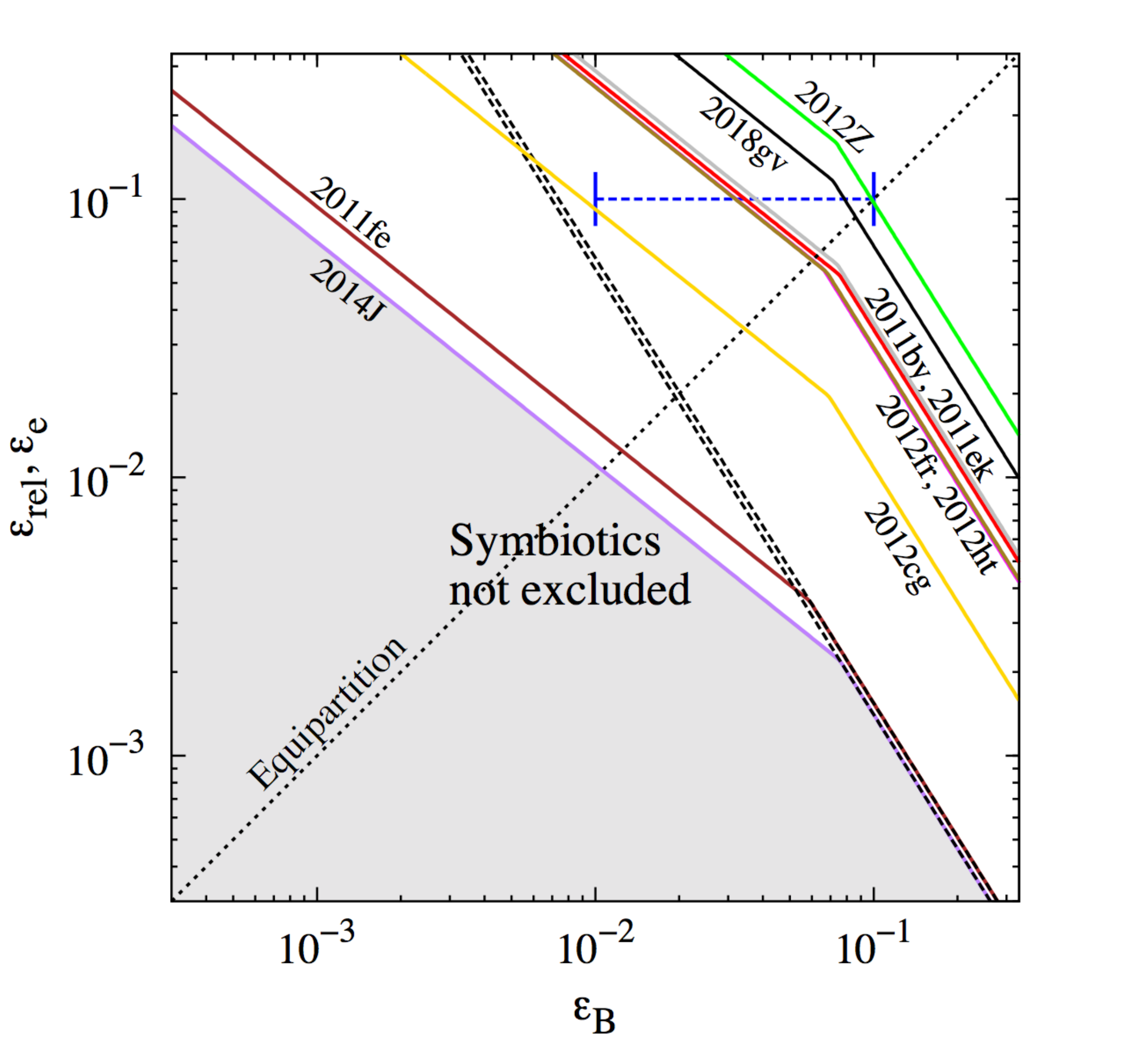}
\caption{
Parameter $\epsilon_{\rm B}$ versus parameters $\epsilon_{\rm e}$ and
$\epsilon_{\rm rel}$ for the nine SNe~Ia with the lowest estimated
$\mdot/v_{w}$ in Table~\ref{tab:RadioLog3}. The curve for
each SN shows the combination of $\epsilon_{\rm B}$ and
$\epsilon_{\rm rel}$ which gives an estimated $\mdot/v_{w}$
corresponding to the upper left corner in the `Symbiotics' box in
Figure~\ref{fig:mdot-vwind}. For combinations of $\epsilon_{\rm B}$ 
and $\epsilon_{\rm rel}$ below, and to the left of these curves,
symbiotic progenitor systems cannot be excluded. The grey area shows 
this parameter space for SN~2014J. The dashed lines for SNe~2011fe 
and 2014J show $\epsilon_{\rm B}$ versus $\epsilon_{\rm e}$ 
corresponding to the $\epsilon_{\rm B}$ vs. $\epsilon_{\rm rel}$
solutions for the these SNe. Supernovae towards the upper right 
corner are progressively less constraining with regard to symbiotics
as a viable progenitor scenario. The blue dashed line depicts the
range for $\epsilon_{\rm B}$ and $\epsilon_{\rm rel}$ normally used 
in models for radio emission from SNe~Ia, i.e., $0.01-0.1$ for 
$\epsilon_{\rm B}$ and $\epsilon_{\rm rel} = 0.1$ \citep[e.g.,][]{cho12,cho16,per14}. As can be seen, for this interval, symbiotics can be excluded only for SNe~2011fe, 2012cg and 2014J. See text for further details.
}
\label{fig:symb}
\end{figure}

In Figure~\ref{fig:symb}, we show solutions for several of the
SNe in Table \ref{tab:RadioLog3}. The figure shows
solutions for $\epsilon_{\rm B}$ and $\epsilon_{\rm rel}$ for 
an assumed wind described by 
$\dot{M} = 1.7 \EE{-8} \msunyr (v_w/100 \kms)$,
which corresponds to the upper left corner of the `Symbiotics'
box in Figure~\ref{fig:mdot-vwind}. For combinations of 
$\epsilon_{\rm B}$ and $\epsilon_{\rm rel}$ lying below, and to the
left, of the solution curves, a symbiotic progenitor system
cannot be excluded, based on the radio data in
Table~\ref{tab:RadioLog3} alone.
For our standard set of model parameters (i.e., $n=13$, $s=2$, 
$T_{\rm bright} = 5\EE{10}$~K, and the N100 explosion model)
$\epsilon_{\rm rel} = \epsilon_{\rm e}$ in Figure~\ref{fig:symb} when 
$\epsilon_{\rm B} \gsim 0.1$, whereas for lower values of
$\epsilon_{\rm B}$, $\epsilon_{\rm rel} < \epsilon_{\rm e}$ 
(cf. Section~\ref{sec:model}). 
The vertical axis of the figure can also be used for $\epsilon_{\rm e}$ if one 
makes extrapolations toward smaller values of $\epsilon_{\rm B}$, like those 
extrapolations shown by dashed black lines for SNe~2011fe 
and 2014J. 

The horizontal blue dashed line highlights 
$0.01 \leq \epsilon_{\rm B} \leq 0.1$ for $\epsilon_{\rm rel} =0.1$
In most analyses, only this small stretch in the  $\epsilon_{\rm B} - \epsilon_{\rm rel}$ plane 
is explored \citep[e.g.,][]{cho12,cho16,per14,kun17}, and, except in a few cases
\citep[e.g.,][]{kun17}, $\epsilon_{\rm rel}$ is allowed to deviate from $\epsilon_{\rm e}$.
Only solutions for SNe~2011fe, 2012cg and 2014J 
lie (in the case of SN~2012cg, marginally) below this blue region, 
which would mean they cannot stem from symbiotic systems if $\epsilon_{\rm rel} =0.1$
and $\epsilon_{\rm B} > 0.01$. 
However, for $\epsilon_{\rm B} = \epsilon_{\rm rel}$ (i.e., 
relativistic particles and magnetic field strength being
in equipartition), and both being $\lsim 0.01$, symbiotic systems
cannot be fully excluded, even for SN~2011fe and 2014J.

\begin{figure}
\centering
\includegraphics[width=12 cm,angle=0]{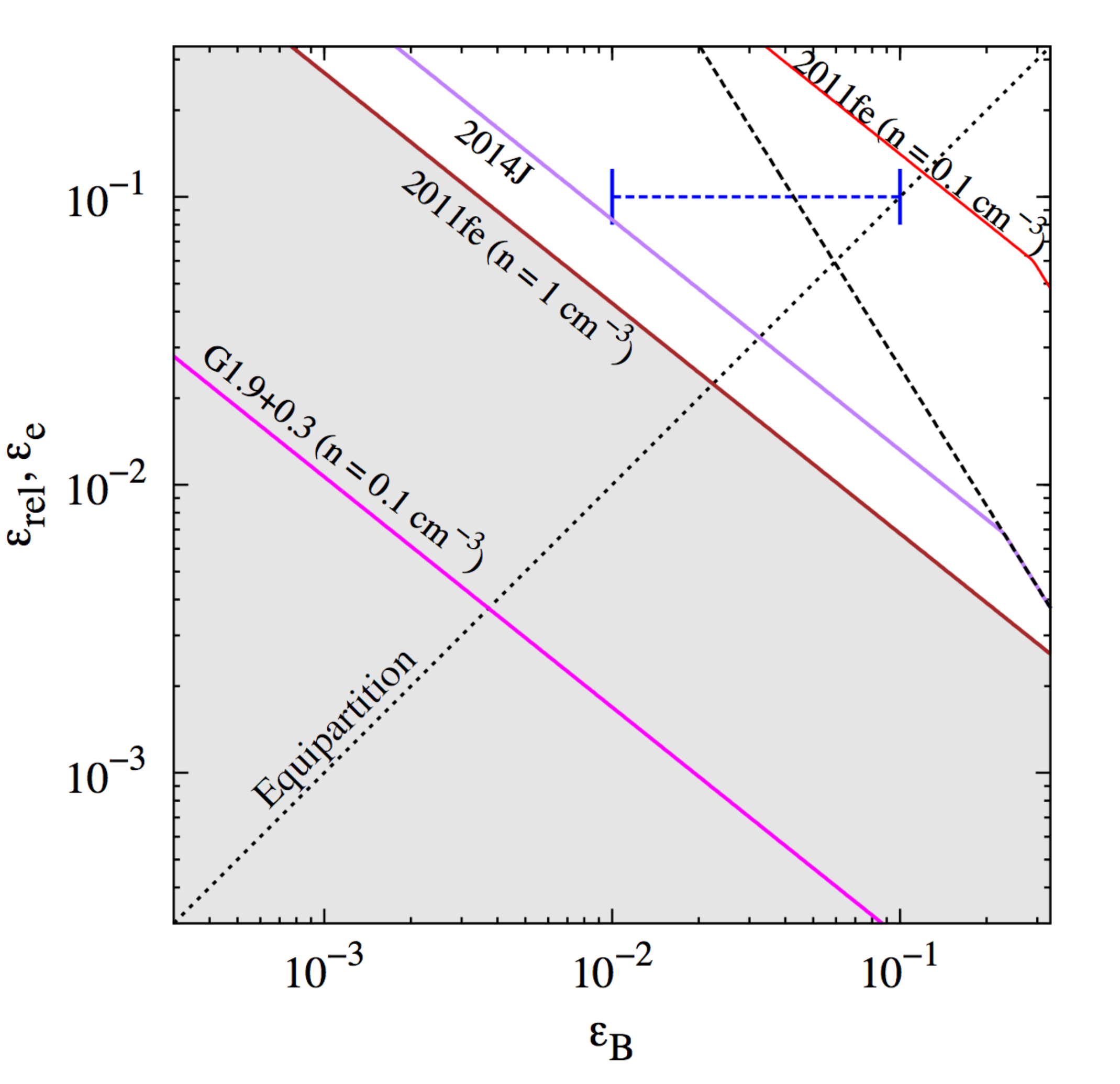}
\caption{
Parameter $\epsilon_{\rm B}$ versus parameters $\epsilon_{\rm e}$ and
$\epsilon_{\rm rel}$ for SNe~2011fe and 2014J, as well as the
young SN~remnant G1.9+03. The curves for SNe~2011fe and 2014J
show the combination of $\epsilon_{\rm B}$ and $\epsilon_{\rm rel}$
corresponding to an upper limit of a constant-density surrounding
medium. For SN~2014J this limit is $n_0 = 1 \cm3$, while for
SN~2011fe we show models for both this density and for 
$n_0 = 0.1 \cm3$. For the detected radio emission from G1.9+03, we
have used $n_0 = 0.1 \cm3$. The dashed black line for
SN~2014J has the same meaning as in Figure~\ref{fig:symb}. 
Data are from \citet{kun17} and
\citet{sar17}. For combinations of $\epsilon_{\rm B}$ 
and $\epsilon_{\rm rel}$ below `2011fe ($n = 1 \cm3$)', a
constant-density medium with $n_0 \geq 1 \cm3$ surrounding
SN~2011fe cannot be excluded from radio observations alone. 
The blue dashed line has the same meaning as in Figure~\ref{fig:symb}.
See text for further details.
}
\label{fig:late}
\end{figure}

In Figure~\ref{fig:late} we show a similar diagram for 
$n=13$, $s=0$, $T_{\rm bright} = 5\EE{10}$~K, and the merger
explosion model. For SNe~2011fe and 2014J we have used the 3 GHz
3$\sigma$ upper limit at 1468 days, and 1.66 GHz 3$\sigma$ upper
limit at 410 days, respectively \citep{kun17}. For G1.9+0.3,
the 1.4 GHz luminosity, at the estimated age of 125 yrs, was used. 
This is $(6.4\pm0.3)\EE{22}$~ergs~s$^{-1}$~Hz$^{-1}$
\citep{sar17}. An interesting note is that a remnant elsewhere 
with such a luminosity could be detected with present-day
instrumentation at distances $\lsim 2.3$ Mpc, assuming a 3$\sigma$
upper limit of $10~\mu$Jy. 

For our $s=0$ models of SN~2011fe, SN~2014J and G1.9+0.3 we 
have assumed densities of
the circumstellar/interstellar medium to be $0.1-1.0~\cm3$, and 
we show the influence on the derived solutions for 
$\epsilon_{\rm B}$ and $\epsilon_{\rm rel}$ for this range in 
$n_0$ in Figure~\ref{fig:late} for SN~2011fe. From this it can be seen
that a density as low as $0.1~\cm3$ would require high efficiency 
of magnetic field amplification and creation of relativistic
particle energy density, so that both $\epsilon_{\rm B}$ 
and $\epsilon_{\rm rel}$ would have to be in excess of 0.1 (if
in equipartition) to correspond to the observed radio upper limit.
However, as discussed in \citet{kun17}, both SN~2011fe and 2014J
are likely to have exploded in an interstellar region with 
density $\sim 1~\cm3$. The solution for SN~2014J in
Figure~\ref{fig:late} is for that density. If any of those
SNe would stem from a DD scenario, they would therefore indicate
that the values for $\epsilon_{\rm B}$ and $\epsilon_{\rm rel}$ 
could be smaller than the standard range marked by the blue region 
in Figure~\ref{fig:late}. 

A caveat with the model run for SN~2011fe
exploding into a $\sim 1~\cm3$ environment, is that our 
assumption of $n=13$ only holds for maximum ejecta velocities
of $\gsim 2.5\EE{9}~\cms$. At lower velocities, the ejecta slope
gets flatter \citep{kun17}. A careful check shows that the
maximum ejecta velocity is $\approx 3.05\EE{9}~\cms$ at 1468 
days for $n=13$ and $n_0 = 1~\cm3$, so the solution for 
SN~2011fe in Figure~\ref{fig:late} is not outside model bounds.
However, for G1.9+0.3, with $n_0 = 0.1~\cm3$ (as in
Figure~\ref{fig:late}), $v_s \sim 1.45\EE{9} \cms$ (and the 
maximum ejecta velocity is $\approx 1.65\EE{9}~\cms$) at 
125 years, which is at the base of the steep outer ejecta
\citep[cf. Fig. 1 of][]{kun17}. The observed velocities of the
expanding radio structures are actually $\lsim 10^9~\cms$
\citep{sar17}, which means the reverse shock has advanced deeper 
into the ejecta than in our model. Moreover, $T_{\rm bright}$ is
unlikely to remain constant over such a long period. Our model
for G1.9+0.3 should therefore only serve as rough estimates
for $\epsilon_{\rm B}$ and $\epsilon_{\rm rel}$.
With this in mind, for $\epsilon_{\rm B} = 0.02$ in
Figure~\ref{fig:late}
we obtain $\epsilon_{\rm rel} \sim 0.003$. In their models, more
tuned to the remnant stage, \citet{sar17} use 
$\epsilon_{\rm rel} = 10^{-4}$ and $p=2.2$ to obtain a best fit 
for $n_0 = 0.18~\cm3$.
Although there is some controversy regarding the density around
G1.9+0.3, probably in the range $0.02-0.3~\cm3$ 
\citep[][and references therein]{sar17}, densities much
less than $n_0 \sim 0.1~\cm3$ may confront the apparent slow
propagation of radio structures. It therefore seems reasonable 
to assume that $\epsilon_{\rm B}$ and $\epsilon_{\rm rel}$ are low
for G1.9+0.3, as they also appear to be for slightly older remnants
\citep[e.g.,][]{mar16,sar17}

In summary, both $\epsilon_{\rm B}$ and $\epsilon_{\rm rel}$ are
at present probably too uncertain to exclude most SD scenarios in 
Figure~\ref{fig:mdot-vwind}.  If we use SN~2011dh as an example to constrain
microphysics parameters for SNe~Ia, we note that  \citet{sod12} 
argue for $\epsilon_{\rm B} = 0.01$ and $\epsilon_{\rm e} / \epsilon_{\rm B} \approx 30$
for that SN. \citet{kun18} estimate a factor of $\sim 6.7$.higher circumstellar density 
than \citet{sod12}, and therefore have to invoke less efficient radio production. 
If we assume $\epsilon_{\rm e} / \epsilon_{\rm B} \approx 30$, as
did \citet{sod12}, Equation~\ref{eq:Lum12} and the study of  \citet{kun18} suggest 
$\epsilon_{\rm e} \approx 0.11$ and $\epsilon_{\rm B} \approx 0.0036$ for 
SN~2011dh, rather than $\epsilon_{\rm e} = 0.30$ and $\epsilon_{\rm B} = 0.01$ 
argued for by \citet{sod12} . If we further compensate for $T_{\rm bright} = 5\EE{10}$~K 
used in the analysis here and $T_{\rm bright} = 4\EE{10}$~K used by
\citet{kun18} for SN~2011dh, $\epsilon_{\rm e} \approx 0.1$ and
$\epsilon_{\rm B} \approx 0.0033$ may provide possible parameter
values for SN~Ia. If we use those numbers in a model based on N100 and
$n=13$, the upper limits on mass loss for SNe~2011fe (2014J) are
$\dot{M} \approx 9.8~(9.1)\times 10^{-9}\,M_{\odot}$\ yr$^{-1}$ (for 
$v_w = 100~\kms$), and for SN~2012cg it is $\dot{M} \approx 3.9 \times 10^{-8}\,M_{\odot}$\ yr$^{-1}$.
This means that such a combination of
$\epsilon_{\rm rel}$ and $\epsilon_{\rm B}$ would 
fully rule out symbiotics for SNe~2011fe and 2014J, 
but not for any of the other SNe in Table~\ref{tab:RadioLog3}. 

The uncertainty in especially $\epsilon_{\rm B}$ is not
surprising from a theoretical point of view. Our current
understanding of shock formation suggests the creation of intense
turbulence with $\epsilon_{\rm B} \sim 0.01$ immediately behind 
the shock \citep{mar16}, but how this high level of turbulence
can be maintained throughout the post-shock region is a conundrum.
It may in fact be that the generation of magnetic field energy 
density is mainly driven by large scale instabilities in
connection with the contact discontinuity. If so, $\epsilon_{\rm B}$
would depend less on the conditions at the blast wave than
on, e.g., the structure of the SN ejecta being
overrun by the reverse shock \citep[][]{bjo17}. Spatially resolved
studies, and modeling thereof, of young SNe like SN~1993J and 
young SNRs are essential to constrain this alternative.

\subsection{Other clues to the origin of SNe~Ia}
\label{sec:other}
In addition to radio emission, there are other clues to the origin of 
SNe~Ia. Many of them involve circumstellar matter. We now discuss 
this, with emphasis on the SNe in Table~\ref{tab:RadioLog3}. 

\subsubsection{Circumstellar absorption line features}
\label{sec:other06X}
Among the SNe in Table~\ref{tab:RadioLog3}, SN~2006X shows
the clearest indication of a circumstellar medium, as it 
displayed time variable narrow Na~I~D absorption features
along the line-of-sight to the supernova, at a distance of 
$10^{16} - 10^{17}$~cm from the progenitor system \citep{pat07}. 
In our models
with N100, $\dot{M} = 1\times 10^{-8}\,M_{\odot}$\ yr$^{-1}$
(and $v_w = 100~\kms$), and $n=13$, interaction with such a shell
would start between $17-215$ days after the explosion. Radio
observations of the supernova, unfortunately, had a gap between
days $18-287$ \citep{cho16}, so any temporary radio increase
could have been missed, especially if the shell had
modest thickness \citep[cf.][who constructed models for radio emission in
shell-like media]{har16}. 

The presumed shell around SN~2006X would signal an SD scenario, 
but it does not have to be the result of a shell ejection. 
It could also exist in the so-called spun-up/spun-down
super-Chandrasekhar mass WD scenario \citep{dis11,jus11,hachisu12}.
Here the donor star shrinks far inside its Roche lobe prior to the
explosion, and dilute circumstellar gas, with density similar
to interstallar gas, would be expected close to the WD.  
If Roche-lobe overflow ceased some $\sim 10^3$ years ago, 
and the wind speed of the non-conservative mass-loss was 
$100 \kms$, dense circumstellar gas could reside at a distance 
of $\sim 3\EE{17}$ cm, and may explain the presumed shell 
around SN~2006X. A shell at such a distance from the supernova
would not be reached by the blast wave until after almost three 
years (using N100, $n=13$ and $n_0=1~\cm3$ inside the shell.) 
This is much later than the last radio observation of SN~2006X,
performed on day 290, indicating a circumstellar density of 
$n_0 \lesssim 50~\cm3$, assuming $\epsilon_{\rm B} = 
\epsilon_{\rm rel} = 0.1$ \citep{cho16}.

Pinning down a possible increase in circumstellar density at
$10^{17} - 10^{18}$~cm from the supernova was one of the
motivations for the late epoch observations of SNe~2011fe and
2014J presented by \citet{kun17}. While SN~2011fe showed no
obvious evidence of circumstellar shells \citep{pat13}, SN~2014J
indeed displayed variations in narrow absorption of
K~I~$\lambda$7665 \citep{gra15}. However, the absorbing gas is 
at $\sim 10^{19}$~cm, and is of interstellar origin \citep{mae16}. 
As described in \citet{kun17}, no radio emission at late epoch 
was detected for SN~2014J, limiting the estimated circumstellar density to 
$n_0 \lesssim 0.4~\cm3$, assuming $\epsilon_{\rm B} = 
\epsilon_{\rm rel} = 0.1$ \citep{kun17}. For such a circumstellar
density, it would take $\gsim 200$ yrs for the SN ejecta, using
the N100 model, to reach a shell at $10^{19}$~cm, i.e., the
supernova would then be in the SNR stage.

\subsubsection{Circumstellar emission and interaction}
\label{sec:circemission}
A small fraction of SNe~Ia show intense circumstellar interaction (cf. Section~1)
and Balmer line emission. \citet{gra18} estimate that probably significantly less 
than $<6$\% \citep{gra18} have circumstellar shells within $< 3\EE{17}$~cm from the 
exploding star giving rise to such emission. The mass of these
shells can be large, perhaps several solar masses \citep[e.g.,][]{ham03,ald06}.
This has been interpreted as clear evidence of SD progenitor systems for at
least this fraction of SNe~Ia. 

As described in Sections~\ref{sec:constant} and \ref{sec:other06X},
very few SNe~Ia have been observed at depth at late epochs to possibly 
detect radio emission resulting from circumstellar interaction. Among those
SN~2006X, and now recently SN~2015cp \citep{har18}, show evidence of circumstellar 
interaction from observations at other wavelengths. As discussed in Section~\ref{sec:other06X}, 
the timing of the radio observations of SN~2006X may have been unfortunate; the 
importance of continuous radio monitoring of SNe~Ia with circumstellar interaction
was discussed by \citet{chu04} for SN~2001ic, as well as by \citet{har18} for SN~2015cp. 
\citet{har16,har18} model how the distribution of the circumstellar gas affects the
expected radio emission.  

An immediate method to probe circumstellar gas is
through X-ray observations, and the only SN~Ia detected in X-rays is SN~2012ca
\citep{boch17}. This supernova belongs to the class of SN~Ia H$\alpha$ emitters,
and the mass of the circumstellar shell is at least $0.1\pm0.05~\msun$. The relaltive
proximity ($\sim 80$~Mpc) of SN~2012ca compared to, e.g., SNe~2002ic and 2005gj, 
is consistent with SN~2012ca being detected in X-rays, and the other two not \citep[cf.][]{hug07}.

\subsubsection{X-ray observations of SN~2011fe, 2012cg and SN~2014J}
\label{sec:other11fe14J}
\citet{mar12,mar14} provided 
deep X-ray limits for SNe~2011fe and 2014J. The X-ray emission
is for early epochs supposed to be due to inverse Compton scattering
of photospheric photons on relativistic electrons in the shocked
circumstellar gas. The derived limits on wind density do not 
depend on $\epsilon_{\rm B}$, but has an $\epsilon_{\rm rel}^{-2}$ dependence.
\citet{mar12} find $\dot{M} \lesssim 2\EE{-9}~(v_w/100~ \kms)~(\epsilon_{\rm rel}/0.1)^{-2}~\msunyr$ for SN~2011fe, and
\citet{mar14} find $\dot{M} \lesssim 1.2\EE{-9}~(v_w/100~ \kms)~(\epsilon_{\rm rel}/0.1)^{-2}~\msunyr$ for SN~2014J.
We can combine this with Equation~\ref{eq:Lum12}, and entries in
Table~\ref{tab:RadioLog3} for SNe~2011fe and 2014J, to get
$\epsilon_{\rm B} \lesssim 0.03~(0.06)$ for SN~2011fe (SN~2014J)
for the X-ray upper limit to be stricter than the radio limit, assuming 
$\epsilon_{\rm e} = \epsilon_{\rm rel} = 0.1$ (where the first
equality holds early in the evolution, i.e., when the most 
constraining X-ray observations where performed for these SNe).
For larger values of $\epsilon_{\rm B}$, radio is more
constraining than X-rays.

Recently, X-ray observations have also been reported for
SN~2012cg \citep{sha18}, and the absence of X-ray emission is claimed to provide 
an upper limit on $\dot{M}$ which is
$\dot{M} \lesssim 1\EE{-6}~(v_w/100~ \kms)~(\epsilon_{\rm rel}/0.1)^{-2}~\msunyr$. In the model used by \citet{sha18}
$\epsilon_{\rm rel}$ is forced to have the same value as 
$\epsilon_{\rm e}$. However, at such high values of
$\dot{M}$, our simulations with N100, $n=13$ and $t=5$ days 
gives ($\epsilon_{\rm rel}/0.1)^{2} \simeq 0.27$ for $\dot{M} = 4\EE{-6}~(v_w/(100~ \kms))~\msunyr$, which should 
be a more correct upper limit of $\dot{M}/v_w$ from the absence
of detected X-ray emission from SN~2012cg. 

The estimated limit
on X-ray luminosity from SN~2012cg, 
$L_{3-10~{\rm keV}} < 1.4\EE{39}$~erg~s$^{-1}$, is too high to 
be in conflict with the expected thermal X-ray emission for 
$\dot{M} = 4\EE{-6}~(v_w/(100~ \kms))~\msunyr$  \citep{lun13}, but 
such a high mass loss rate would have repercussions for
interpretations of the the radio data. While FFA is below unity
($\tau_{\rm ff}\sim 0.08$, see 
Section~\ref{sec:modelingsample}) for the 4.1 GHz observations at 
5 days (cf. Table~\ref{tab:RadioLog3}), SSA would make the radio 
flux not peak until after $\sim 55$~days at 4.1 GHz,
if $\epsilon_{\rm e} = \epsilon_{\rm rel} = 0.1$.
Despite SSA, the luminosity at 5 days is much higher than
listed in Table~\ref{tab:RadioLog3}. In order not to violate the
observed 4.1 GHz flux, $\epsilon_{\rm B} \lsim 3\EE{-6}$, 
assuming $\epsilon_{\rm e} = 0.1$, and other model parameters for 
our $n=13$ and $s=2$ simulations.
A combination of $\epsilon_{\rm rel} = 0.027$ and 
$\epsilon_{\rm B} \lsim 3\EE{-6}$ is probably extreme, and it
is therefore most likely safe to assume that the X-ray observations
of SN~2012cg are much less constraining than the radio data for 
this SN in terms of $\dot{M}/v_w$.

\subsubsection{Dust extinction of SN~1989B, SN~2006X, SN~2012cg, SN~2012cu and SN~2014J}
\label{sec:other12cg}
Circumstellar matter may reveal its presence through dust
signatures. Among the SNe in Table~\ref{tab:RadioLog3}, SNe~2012cg, 2012cu and 2014J were investigated by \citet{ram15}
to look for extinction features that could be due to
circumstellar matter. For SNe~2012cu and 2014J, no color evolution
of the extinction was found, while for SN~2012cg there is evidence
of some evolution. This could argue for circumstellar dust in 
SN~2012cg. However, when complementing with high-resolution data 
of Na~I~D, \citet{ram15} argue that any such dust around 
SN~2012cg must be at a distance of $\gsim 10^{19}$~cm, which 
does not necessarily relate it to the progenitor system. The
density probed by the published latest radio data, i.e., 
at 216 days, gives $n_0 \lesssim 10~\cm3$, assuming 
$\epsilon_{\rm B} = \epsilon_{\rm rel} = 0.1$ \citep{cho16}. 
Using the N100 model with $n=13$ and $n_0 = 10~\cm3$, the 
blast wave had only expanded out to $\simeq8\EE{16}$~cm
at that epoch, i.e., far inside the minimum distance to the
dust.

In a more recent dust study, \citet{bul18} analyze 48 reddened
SNe~Ia in order to localize sources of dust extinction. Supernovae
appearing in both that study, and in Table~\ref{tab:RadioLog3},
are: SNe~1989B, 2006X, 2012cu and 2014J. From the models of \citet{bul18}, 
the distance between supernova and dust
for SNe~1989B and 2012cu is $\gsim 4.3\EE{19}$~cm and
$\gsim 1.0\EE{19}$~cm, respectively. For SNe~2006X and 2014J, the
dust is mainly located $\sim 5\EE{19}$~cm and $\sim 1.4\EE{20}$~cm
from the supernova, respectively. Only one supernova in the study 
of \citet{bul18}, namely SN~2003hx, has dust close enough to the
supernova, $\sim 4\EE{16}$~cm, to argue for it being
circumstellar. However, this supernova is close to the center 
of its host galaxy, and \citet{bul18} conclude that neither this, nor 
any of the other SN~Ia in their study, should be considered to
harbor circumstellar dust. The dust is likely interstellar in all
their cases.

Comparing with SNe in Table~\ref{tab:RadioLog3}, we note that 
SN~2012cu was observed only once in radio, while
SN~1989B was monitored until 114 days after the explosion.
\citet{cho16} estimate $n_0 \lesssim 40~\cm3$ for that epoch, 
assuming $\epsilon_{\rm B} = \epsilon_{\rm rel} = 0.1$.

\subsubsection{Interaction with a binary companion}
\label{sec:collision}
In the SD scenario, the donor will be overrun by the supernova blast-wave
in $\sim 0.6~(v_s / 5\EE{4}~\kms)^{-1}~(R_{\rm sep}/10^{13}~{\rm cm})$
hours, where $R_{\rm sep}$ is the separation between the donor and the
WD at the time of explosion. The donor will therefore quickly be 
hidden inside the SN ejecta. However, during this early phase, and shortly 
thereafter, the donor can give rise to observational signatures in X-rays and
optical/UV, strength depending on the viewing angle \citep{kas10}. Caught early
enough, $\sim 10\%$ of SD cases should give rise to detectable signatures. 
In general, early interaction may create a light curve that would deviate from a 
single power-law. Such cases have indeed been identified, e.g., 
SN~2012fr \citep{con18}, SN~2013dy \citep{zheng13}, 
SN~2014J \citep{goo15,siv15}, MUSSES1604D \citep{jiang17}, iPTF16abc \citep{mill18},
SN 2017cbv \citep{hoss17} and ASASSN-18bt \citep{shapp18a}. 
However, searches for other markers of SD origin have proven negative. 

Of particular
interest here are SNe~2012fr and 2014J which both are among the SNe~Ia
with the most constraining radio limits on circumstellar matter and microphysicsin 
parameter in the SD scenario (cf. Table~\ref{tab:RadioLog3} and Figure~\ref{fig:symb}).
This could signal that the early light curve behavior is caused by something else 
than ejecta-companion interaction.

A hint to another origin is the finding by \citet{stri18} that there are two well-defined classes of
SNe~Ia, one of which has a blue color for the first few days, and the other a red color. In addition, 
there is a correlation between the early blue color and photospheric temperature at maximum light.
At maximum, a SD companion should be well hidden by the supernova ejecta, and the SN light
is powered by radioactive decay. It is not clear why this should correlate with early blue color
resulting from ejecta-companion interaction. Further statistics is needed to shed light on this.

\subsubsection{Nebular emission}
\label{sec:nebular}
Long after the initial phases discussed in Section~\ref{sec:collision}, a SD
scenario donor may potentially reveal itself, but not until the optical
depth through the ejecta has dropped for the donor material to become visible. 
In the 1D models of \citet{mat05} and \citet{lun13}, this was calculated
to occur after a few hundred days. In particular, lines of
hydrogen, or perhaps helium, calcium or oxygen \citep{lun15},
with an expected velocity width of $\sim (0.5-2)\times 10^3~\kms$
\citep[e.g.,][]{liu12,liu13,pan12,boe17} would indicate an SD scenario.
The estimated amount of ablated gas from the donor varies depending
on donor size and type, and separation between the donor and the WD, 
but typical values are $\sim 0.01-0.1~\msun$.

Several studies have been done in the nebular phase of SNe~Ia
to look for material from a putative non-degenerate companion,
using the models of \citet{mat05} and \citet{lun13}, and
in many cases the estimated upper limit of hydrogen mass from
the companion is $\lesssim 0.01~\msun$ 
\citep[e.g.,][]{leo07,sha13, sha18, lun15,mag16}. 
For our sample in
Table~\ref{tab:RadioLog3}, SNe~2011ek, 2011fe, 2011iv, 2012cg,
2012cu, 2012fr, 2012ht and 2014J have all been studied in the 
nebular phase, and the mass of hydrogen-rich donor material
is $\lsim 0.01~\msun$, except for SN~2012cu, for
which the limit is higher.

The most recent models for the expected emission from donor
material in the nebular phase \citep{boe17,sand18,dim18,tuck18}
indicate that the mass limits on ablated gas could be even lower
than those derived from the models of \citet{mat05} and \citet{lun13}. 
However, in all studies, systematic errors in the 
mass estimates could have been underestimated \citep{lun15}, 
as the underlying SN spectrum can have intrinsic spectral features \citep[e.g.,][]{bla18} 
that may mask emission from ablated donor material. Time sequences of nebular 
spectra are needed to remove this uncertainty, as well as confusion due to
other excitation mechanisms than radioactivity. This is highlighted
by the claimed detection of ablated material in ASASSN-18tb (SN~2018fhw) \citep{Koll19} at 
a single epoch of around $155-160$ days pasty explosion. A sequence of spectra
of this event shows a persistent H$\alpha$ emission already $\lsim 60$ days \citep{Val19}
after explosion, which is more the hallmark of circumstellar interaction.

Despite some remaining uncertainty, the mass limits on ablated 
donor material from the absence of the nebular emission lines discussed
in  \citet{lun15} are in conflict with the hydrodynamic models 
of the WD-companion interaction, and pose a serious challenge
to SD scenarios. The only possible SD scenario surviving this
observational test may in fact be
that of a spun-up/spun-down super-Chandrasekhar mass donor 
\citep[see][for a discussion on this]{lun15}. From a
circumstellar point of view, this would suggest that
a constant circumstellar density out to some radius,
corresponding to when mass transfer from the donor ceased
(cf. Section \ref{sec:other06X}), may provide the most likely
circmstellar structure in the SD scenario. A low-density medium is
also expected in the DD scenario. It may therefore not be
surprising that SNe~Ia are still undetected in radio.

\subsubsection{SN~2013dy, SN~2016coj, SN~2018gv, SN~2018pv and SN~2019np}
\label{sec:other13dy}
There is no evidence of circumstellar material in any of 
SNe~2013dy, 2016coj, 2018gv, 2018pv and 2019np. The most well-studied 
of them is SN~2013dy \citep{zheng13,pan15,zhai16}. 
It was detected $\sim 2.4$ hours after first light, and had 
an abundance of unburned material in its envelope. B-band 
maximum occurred after $\sim 17.7$ days, and our radio 
observation was made $\sim 8$ days later. High-resolution 
optical spectra were obtained by \citet{pan15}, but no
variability was found in the standard absorption lines 
Ca H\&K, Na D~$\lambda\lambda$5890,~5896 and 
K~I~$\lambda\lambda$7665, 7699. These authors also show
nebular spectra until 333 days after maximum, but no mass 
limits on possible donor material were presented. 

SN~2016coj is estimated to have been detected $\sim 4.9$ days
after first light \citep{zheng17}. It is a spectroscopically 
normal SN, with a B-band maximum at $\sim 16$ days. 
High-resolution spectra were obtained \citep{zheng17}, but 
owing to its $\sim 20$ Mpc distance, the S/N
ratio was too low to identify any interstellar or circumstellar 
lines. SN~2016coj is the supernova in our sample with the 
largest number of radio observations. They are, however, not as
deep as for SNe~2018gv, 2018pv and 2019np. 

Observations of SNe~2018gv and 2018pv at other
wavelengths than radio are discussed by
P. Chen et al. (in preparation). For SN~2018pv
we have included ASAS-SN data
\citep[see][for a description of ASAS-SN]{shappee14}.  B-band 
data for SN~2019np have been estimated from data made available
by N. Elias-Rosa and S. Dong.
In Table~\ref{tab:RadioLog2} we have entered 6 days since explosion
for SN~2018gv. This is a conservative estimate. From the optical data we
have consulted, it may be closer to 5 days. This would push 
$\dot{M}/v_w$ close to $1.1\EE{-8}~(v_w/100~ \kms)~\msunyr$. Likewise, the 14 
and 10 days entered for SNe~2018pv and 2019np are conservative upper limits on 
the time since explosion. 

Regarding other tests for the SD
scenario, SN~2018gv with its small host confusion, should be 
an excellent target for nebular emission studies. This could
test the suggestion by \citet{yang19} that this supernova 
was indeed a member of a SD progenitor system, based on early ($-13.6$ days with respect to 
B-band maximum light) spectropolarimetric measurements. The supernova showed only 
$\leq 0.2$\% continuum polarization, as well as moderate line polarization, $0.46 \pm 0.04$\%, for 
the strong \ion{Si}{2}~$\lambda$6355 and $0.88 \pm 0.04$\% for the high-velocity \ion{Ca}{2} component. 
\citet{yang19} claims that this is inconsistent with a DD scenario. This is not in conflict 
with our radio limits (cf. Figure~\ref{fig:mdot-vwind}), if the progenitor was part of
a symbiotic system, and/or at least one of $\epsilon_{\rm e}$ and $\epsilon_{\rm B}$ had 
a value $< 0.1$. 

\subsection{Future radio observations}
\label{sec:future}
\begin{figure}
\centering
\includegraphics[width=12 cm,angle=0]{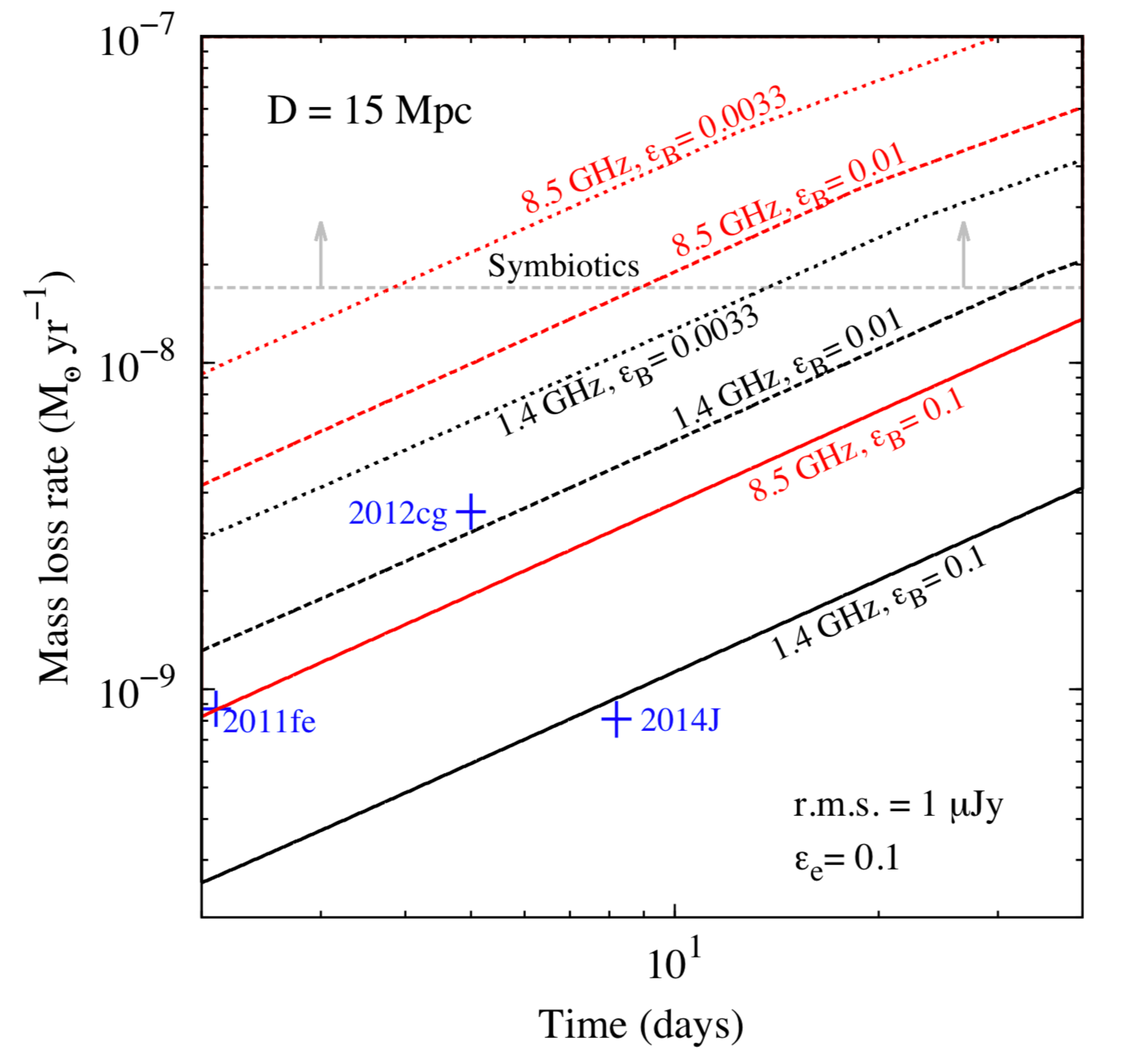}
\caption{
Same as Figure~\ref{fig:Time_Mdot}, but for the fixed r.m.s. of $1.0~\mu$Jy
expected to be the sensitivity of SKA1-mid phase.
Upper limits on mass loss rate for two frequencies (1.4 GHz and 8.5 GHz) are drawn for 
3$\sigma$ flux limits. For each frequency solutions are made for three different values of
$\epsilon_{\rm B}$, namely 0.1 (solid lines), 0.01 (dashed lines) and 0.0033 (dotted
lines). $\epsilon_{\rm e} = 0.1$ for all models. Mass-loss rate limits for the hitherto most 
constraining events, SNe~2011fe, 2012cg and 2014J, are marked in blue, assuming
$\epsilon_{\rm B} = \epsilon_{\rm e} = 0.1$. The expected lowest mass-loss rate for 
symbiotic systems are marked in gray (cf. Figure~\ref{fig:mdot-vwind}).
See text for further details. 
}
\label{fig:SKA}
\end{figure}

The deepest radio limits on circumstellar gas are for 
SNe~2011fe and 2014J. A leap in sensitivity will occur when
the Square Kilometre Array (SKA) comes online. In the SKA1-mid
phase, a 1$\sigma$ sensitivity of $\sim 1.0~\mu$Jy/beam can be
reached in a one-hour integration at 1.4 GHz. The same limit is also expected at
higher frequencies (e.g., 8.5 GHz and 15 GHz). In Figure~\ref{fig:SKA} we
show a plot similar to that in Figure~\ref{fig:Time_Mdot}, but tuned to detection
limits more relevant for SKA.

Judging from Figure~\ref{fig:SKA}, such a limit at 1.4 GHz will probe 
$\dot{M}/v_w$ down to $\sim 3.7\EE{-10}~(v_w/100~ \kms)~\msunyr$
for a SN at 15 Mpc, observed 3 days after explosion, assuming the
same model parameters used in Table~\ref{tab:RadioLog3}. (The choice of 
3 days after explosion in the estimate above rests on the 
currently planned cadence of the Large Synoptic Survey Telescope,
LSST, which, like SKA, will be a southern hemisphere facility.) 
This is $\gsim 2.2-2.4$ times lower in $\dot{M}/v_w$ compared to the limits 
for SNe~2011fe and 2014J in Table~\ref{tab:RadioLog3}, despite those SNe~Ia being
much closer than 15 Mpc. For a SN at distance 20 Mpc, the limit on $\dot{M}/v_w$ would be 
$\sim 5.5\EE{-10}~(v_w/100~ \kms)~\msunyr$. Judging from the VLA campaign 
by \citet{cho16}, mainly during 2011 and 2012 (cf. Table~\ref{tab:RadioLog3}), we can expect
$\sim 3-4$ SNe~Ia per year within that distance, so a sample like 
that in Table~\ref{tab:RadioLog3} could be built in just a couple of
years using SKA, and with limits on $\dot{M}/v_w$ which are almost a 
factor of two better than our current limits for SNe~2011fe and
2014J, and $\sim 6$ times better than for SN~2012cg. 

Alternatively, we will be able to constrain $\epsilon_{\rm B}$ 
and $\epsilon_{\rm rel}$ to unprecedented levels. 
Figure~\ref{fig:SKA} shows solutions not only for for $\epsilon_{\rm B} = 0.1$,
but also for $\epsilon_{\rm B} = 0.0033$ and $\epsilon_{\rm B} = 0.01$
(cf. Section~\ref{sec:eps}). For a SN at 15 Mpc, observed at 1.4 GHz 3 days after 
explosion, one would probe mass-loss rates down to 
$\dot{M}  \sim 4.2\EE{-9}~(v_w/100~ \kms)~\msunyr$, even if $\epsilon_{\rm B} = 0.0033$.
This would roughly correspond to the blue dashed line for SN~2012cg in
Figure~\ref{fig:mdot-vwind}, but for $\epsilon_{\rm B} = 0.0033$ instead of
$\epsilon_{\rm B} = 0.1$. For a distance of 5 Mpc, and $\epsilon_{\rm B} = 0.0033$,
one would probe down to $\dot{M} \sim 9.6\EE{-10}~(v_w/100~ \kms)~\msunyr$, and 
one would essentially be bound to detect a SN~Ia, if of SD origin.
For SNRs, a remnant like G1.9+0.3 would be detected with
SKA1-mid at 3$\sigma$ out to $\sim 4$ Mpc. 

A difference between Figures~\ref{fig:Time_Mdot} and~\ref{fig:SKA} is that
$\dot{M}$ cannot be constrained in Figure~\ref{fig:Time_Mdot} for short times since explosion 
and moderate flux limits, in particular for low frequencies. For deep flux limits like that in 
Figure~\ref{fig:SKA} this is not a problem, as synchrotron self-absorption is unimportant
at the low mass-loss rates to be probed by SKA. 


\section{Conclusions}
\label{sec:conclusions}
We report deep e-MERLIN and ATCA radio observations of the Type Ia SNe (SNe~Ia)~2013dy,
2016coj, 2018gv, 2018pv and 2019np, along with modeling of their radio emission.
We do not detect the SNe. For the modeling we use the explosion model 
N100 \citep{rop12,sei13}, in combination with a density $\rho \propto r^{-13}$ for 
the outermost supernova ejecta.  For the microphysical parameters 
$\epsilon_{\rm e}$ and $\epsilon_{\rm B}$ (which are the fractions of energy density of 
the shocked gas going into electrons with a power-law energy distribution and 
magnetic field strength, respectively) we first make the standard
assumption $\epsilon_{\rm e} = \epsilon_{\rm B} = 0.1$.
Often it is assumed that $\epsilon_{\rm e} = \epsilon_{\rm rel}$, where $\epsilon_{\rm rel}$ is 
the fraction of energy going into electrons with $\gamma_{\rm min} \gsim 1$. 
Following \citet{kun17}, we have relaxed that assumption in our models.
With these considerations, we arrive at the upper limit on the mass loss rate  
$\dot{M} \lsim 12~(2.8,1.3, 2.1,1.7) \EE{-8} (v_w/100 \kms)$ \msunyr in a wind scenario, 
for these five SNe, respectively, where $v_w$ is the wind velocity of the mass loss
from the progenitor system.
The limits for SNe~2016coj, 2018gv, 2018pv and 2019np are among the 16 deepest ever. 

We have also compiled data for the 21 SNe~Ia with the lowest limits on $\dot{M}/v_w$ 
(including SNe~2016coj, 2018gv, 2018pv and 2019np), which from our models, with the same assumptions
as above, all have $\dot{M} \lsim 4 \EE{-8} (v_w/100 \kms) \msunyr$. We compare those limits with the 
expected mass loss rate in different single-degenerate (SD) progenitor scenarios. With 
$\epsilon_{\rm e} = \epsilon_{\rm B} = 0.1$, the most nearby SNe in the sample, SNe~2011fe 
and 2014J, are unlikely to be the results of SD progenitors, unless mass transfer from the 
donor ceased long before the explosion, like in the spun-up/spun-down super-Chandrasekhar 
mass WD senario. Alternatively, they are the results of two white dwarfs merging, the so-called 
double-degenerate (DD) route. The latter is supported by the absence of detected 
X-ray emission. As X-ray emission is expected to be due to inverse Compton scattering on 
relativistic electrons behind the supernova blast wave, limits on $\dot{M}/v_w$ from X-rays 
depend on $\epsilon_{\rm rel}$, but not on $\epsilon_{\rm B}$. Assuming that 
$\epsilon_{\rm e} = 0.1$ and using $\dot{M}/v_w$ from X-ray limits, we obtain 
$\epsilon_{\rm B}  \lsim 0.03~(3\EE{-6}, 0.06)$ for SN~2011fe~(2012cg, 2014J), respectively, 
for the X-ray upper limit to be stricter than the radio limit. The small value for SN~2012cg (which
is the third most well-constrained SN~Ia in radio) originates from a relatively poor X-ray limit on 
$\dot{M}/v_w$, which we have revised upwards by a factor of four to 
$\dot{M} \lsim 4 \EE{-6} (v_w/100 \kms)$ \msunyr. 

We caution that the uncertainty in the microphysical parameters (mainly $\epsilon_{\rm B}$) makes 
limits on $\dot{M}/v_w$ from radio somewhat difficult to judge. To study this we have allowed 
$\epsilon_{\rm rel}$ and $\epsilon_{\rm B}$ to take any plausible values. In particular, we have
tested what is the allowed range in $\epsilon_{\rm rel}$ and $\epsilon_{\rm B}$ for the 21 SNe~Ia 
in our sample, for them not to stem from symbiotic progenitor systems, which we have defined to 
have minimum mass-loss rate of $\dot{M} \gsim 1.7 \EE{-8}(v_w/100 \kms)  \msunyr $. 
Symbiotic systems are those of the likely progenitors with the largest $\dot{M}/v_w$. 
Assuming $\epsilon_{\rm rel} = 0.1$, and judging from radio alone, the progenitors of even SNe~2011fe 
and 2014J could have had such high mass loss rates for $\epsilon_{\rm B} \lsim 10^{-3}$.
However, $\dot{M} \gsim 1.7 \EE{-8} (v_w/100 \kms)  \msunyr $, is
ruled out from X-ray non-detections for these SNe, if $\epsilon_{\rm rel} = 0.1$.  
A combination with relativistic electrons 
and the magnetic field strength in equipartition, so that $\epsilon_{\rm rel} = \epsilon_{\rm B} = 0.01$, 
could make symbiotic progenitors for SNe~2011fe and 2014J with 
$\dot{M} \gsim 1.7 \EE{-8} (v_w/100 \kms)  \msunyr$ pass observational tests in radio and 
X-rays. For SN~2012cg, $\epsilon_{\rm rel} = 0.1$ and $\epsilon_{\rm B} = 0.01$ is enough to
rule out a symbiotic progenitor, while for other SNe in the sample, radio limits cannot rule
out symbiotic progenitor systems, let alone other SD channels with lower mass-loss rates, even
if $\epsilon_{\rm B}$ is as high as $\sim 0.04$ (assuming $\epsilon_{\rm rel} = 0.1$).

To draw conclusions on progenitor origin from radio and X-rays, it is thus imperative to know the 
microphysical parameters. Information can be provided by objects with actual detections. One 
is the youngest SN~Ia remnant detected in radio, G1.9+0.3. Although there is some debate regarding 
the density around G1.9+0.3, its detection at an age of $\sim 125$ years points towards 
$\epsilon_{\rm rel}$ and $\epsilon_{\rm B}$ both being of order 0.01, or less. With such numbers 
for SN~2011fe and 2014J at late epochs (i.e., $t = 1-3$~years), as well as the 21 SNe~Ia in our 
sample at early epochs, it comes as no surprise there is yet no radio detections of 
SNe~Ia, or young SN~Ia remnants, besides local events like G1.9+0.3, and possibly 
SN~1885A \citep{sar17}. 

Estimates of $\epsilon_{\rm rel}$ and $\epsilon_{\rm B}$ can also be obtained from stripped-envelope 
core-collapse SNe. We have highlighted SN~2011dh
as an example, and argue for $\epsilon_{\rm e} \approx 0.1$ and $\epsilon_{\rm B} \approx 0.0033$. 
Such a combination would fully rule out symbiotics for SNe~2011fe and 2014J,
but not for any of the other SNe~Ia.

When radio observations of a newly detected SN~Ia are being planned, it is crucial to take into account 
synchrotron self-absorption (SSA). As we show in Figure~\ref{fig:Time_Mdot}, too early low-frequency 
($\lsim 2$~GHz) observations may lead to no constraints on circumstellar matter if the 3$\sigma$ flux
limit is too high, and/or the observations are being performed too early. SN~2019np serves as an
example, where 1.28~GHz observations at $t=7$~days (3$\sigma$ limit of 57~$\mu$m) gives no
solution for $\dot{M}/v_w$, and 1.51~GHz observations $t=10$~days (3$\sigma$ limit of 66~$\mu$m)
can be used to rule out $1.7\times10^{-8}\msunyr \lsim {\dot M} (v_w/100 \kms)^{-1}  \lsim 2.4\times10^{-7}\msunyr$
(using our standard parameters for the supernova dynamics and $\epsilon_{\rm rel} = \epsilon_{\rm B} = 0.1$). 
To rule out also $\gsim  2.4\times10^{-7}\msunyr$, complementary observations are needed.
Considering SSA will be important when the Square Kilometre Array (SKA) becomes operational, and
if it will be used to observe moderately distant ($\gsim 40-50$~Mpc) SNe~Ia at low frequencies at early epochs. For
more nearby SNe~Ia, SSA is less important for SKA (cf. Figure~\ref{fig:SKA}) and one should in just a few years create 
a significantly better sample than discussed here.

While radio and X-rays are important probes for circumstellar matter, other tools
are also needed to pin down the origin of SNe~Ia, in particular observations in the optical and infrared. 
Current evidence points in favor of DD being responsible for the majority of normal SNe~Ia, with
the strongest evidence, besides no detected radio or X-ray emission, being no circumstellar dust 
in any SN~Ia \citep{bul18}, no trace of donor material  in nebular 
spectra \citep[e.g.][]{lun15,mag16,sand18,tuck18}, and tight constraints on donor size from 
the very early interaction between SN ejecta and a donor  \citep[e.g.,][]{kas10,sha18}. Evidence for 
circumstellar matter in normal SNe~Ia is provided by time-varying absorption lines in a few SNe~Ia \citep[e.g.][]{pat07}, 
and emission lines in one case \citep{gra18}, and observing such SNe in the radio, at moderate cadence, 
may provide the best prospects of detecting radio emission from a SN~Ia in the near future. 


\acknowledgments
We thank Subo Dong for sharing data for SNe~2018gv and 2018pv prior to publication, Stephen Reynolds for
discussions and comments on an early version of the manuscript, as well as Nancy Elias-Rosa for discussions 
on SN~2019np. PL acknowledges support from the Swedish Research Council, and MPT acknowledges financial 
support from the Spanish MCIU through the ``Center of Excellence Severo Ochoa'' award for the Instituto de
Astrof\'isica de Andaluc\'ia (SEV-2017-0709) and through the MINECO grants
AYA2012-38491-C02-02 and AYA2015-63939-C2-1-P.

The electronic Multi-Element Radio Linked Interferometer 
Network (e-MERLIN) is the UK's facility for high resolution radio astronomy
observations, operated by The University of Manchester for the
Science and Technology Facilities Council (STFC). 
The Australia Telescope Compact Array (ATCA) is part of the Australia Telescope National Facility which 
is funded by the Australian Government for operation as a National Facility managed by CSIRO. 
The ATCA data reported here were obtained under Program C1303 (P.I. P. Lundqvist).
This research has made use of the NASA Astrophysics Data System (ADS) Bibliographic Services, 
and the NASA/IPAC Extragalactic Database (NED), which is operated by the Jet Propulsion Laboratory, 
California Institute of Technology, under contract with the National Aeronautics and Space Administration

\software{CASA \citep{mcm07}, AIPS \citep{Wells85}, MIRIAD \citep{sault95}, e-MERLIN pipeline \citep{Argo15}}

\facility{e-MERLIN, ATCA, JVLA, AMI, MeerKAT}.

\end{document}